%% file: draft.tex
\documentclass[a4paper,11pt]{article}
\pdfoutput=1 

\usepackage{jcappub} 

\usepackage[T1]{fontenc} 
\usepackage{mathtools}
\usepackage{siunitx}

\usepackage{enumitem}
\usepackage{siunitx}
\usepackage[utf8x]{inputenc}

\usepackage{aas_macros}
\usepackage{amssymb}
\usepackage{cleveref}
\usepackage{siunitx}
\usepackage{enumerate}
\usepackage[dvipsnames]{xcolor}
\usepackage{amsmath}

\usepackage{graphicx} 
\usepackage{array}

\usepackage{amsthm}

\usepackage{psfrag}
\usepackage{color}
\usepackage{graphicx}
\usepackage{upgreek}
\usepackage{feynmp}
\DeclareMathAlphabet{\mathpzc}{OT1}{pzc}{m}{it}

\usepackage{tikz}
\usetikzlibrary{trees}
\usetikzlibrary{decorations.pathmorphing}
\usetikzlibrary{decorations.markings}
\tikzset{
    photon/.style={decorate, line width=0.15mm, decoration={snake,amplitude=3pt,segment length=8pt}, draw=black},
    wino/.style={draw=redwine},    
    fermion/.style={draw=black, line width=0.2mm, postaction={decorate},
        decoration={markings,mark=at position .55 with {\arrow[draw=black,scale=2,#1]{>}}}},
    scalar/.style={draw=black, dashed,postaction={decorate},
        decoration={markings,mark=at position .55 with {\arrow[draw=black,scale=2,#1]{>}}}},
    scalarline/.style={draw=black, postaction={decorate},
        decoration={markings,mark=at position .55 with {\arrow[draw=black,scale=2,#1]{>}}}},
    scalarline2/.style={draw=black, postaction={decorate} },
    scalar2/.style={draw=black, dashed,postaction={decorate}},
    gluon/.style={decorate, draw=black,
        decoration={coil,amplitude=3pt, segment length=4pt}},
    graviton/.style={decorate, draw=black,
        decoration={zigzag,amplitude=3pt, segment length=4pt}}
}
\tikzstyle{blob}=[circle,
                              minimum size=20,
                              draw=black!80,
                              fill=black!80]   
\tikzstyle{redblob}=[circle,
                              thick,
                              minimum size=0.4cm,
                              draw=red!80,
                              fill=red!60]

\allowdisplaybreaks[4]

\definecolor{darkgreen}{rgb}{0,0.5,0}

\newcommand{\comment}[1]{}

\newcommand{\bseq}{\begin{subequations}}
\newcommand{\eseq}{\end{subequations}}
\newcommand{\be}{\begin{equation}}
\newcommand{\ee}{\end{equation}}

\newcommand{\RN}[1]{%
  \textup{\uppercase\expandafter{\romannumeral#1}}%
}

\newcommand{\beqa}{\begin{eqnarray}}
\newcommand{\eeqa}{\end{eqnarray}}

\title{
      Superfluid
   Effective Field Theory 
\\
   \centering
   \huge
   for dark matter direct detection  
}

\author{\;Konstantin Matchev,}
\author{Jordan Smolinsky,}
\author{Wei Xue,  \\} 
\author{and Yining You}
\affiliation{Department of Physics, University of Florida,
  Gainesville, FL 32611, USA}

\emailAdd{matchev@ufl.edu}
\emailAdd{jsmolinsky@ufl.edu}
\emailAdd{weixue@ufl.edu}
\emailAdd{youy@ufl.edu}
\abstract{
We develop an effective field theory (EFT) framework for superfluid ${}^4$He to model the interactions among
quasiparticles, helium atoms and probe particles. Our effective field theory approach brings together symmetry arguments and power-counting and matches to classical fluid dynamics. 
We then present the decay and scattering rates for the relevant processes involving quasiparticles and helium atoms. 
The presented EFT framework and results can be used to understand the dynamics of thermalization in the superfluid, and can be further applied to 
sub-GeV dark matter direct detection with superfluid ${}^4$He.
}

\begin{document}
\maketitle
\flushbottom



\section{Introduction}
\label{sec:intro}

\subsection*{Motivation from the dark matter problem}
 
Dark matter (DM) is arguably the biggest mystery in science today. Evidence for it has been steadily accumulating since the 1930's ~\cite{Zwicky:1933gu} and by now its existence at all scales in the Universe is unquestionable~\cite{Bertone:2016nfn,Arbey:2021gdg}. We know it makes up 85\% of the matter in the Universe, yet we remain totally ignorant about its nature. Within the currently accepted and well tested Standard Model of particle physics, none of the known particles is consistent with the profile for a neutral, massive, and stable dark matter particle required to explain the dark matter puzzle. Therefore, the resolution of the dark matter problem necessarily requires \emph{new particles} beyond the Standard Model of particle physics.

Dark matter searches fall into three complementary categories \cite{Bauer:2013ihz}: \emph{collider experiments} which might produce DM particles either directly or in the decays of other, heavier new particles; \emph{indirect detection} of DM annihilation or decay products; and \emph{direct detection} of ambient DM impinging upon instrumented detectors. Current dark matter direct detection experiments are most sensitive to relatively heavy dark matter candidates because they use heavy target materials such as silicon, germanium, and xenon \cite{essig2012direct,essig2012first,essig2016direct,graham2012semiconductor,hochberg2016superconducting,hochberg2016detecting,capparelli2015directional,cavoto2016wimp,bloch2017searching,derenzo2017direct,hochberg2017directional,cavoto2018sub,baracchini2018ptolemy,baxter2020electron,aprile2012dark,akerib2016improved,agnese2014search,tan2016dark,agnese2015wimp,angloher2016results,lee2015modulation}. As a result, there are strong existing limits on the range of dark matter masses above roughly 1 GeV. Furthermore, recent theoretical developments \cite{alexander2016dark,battaglieri2017us,lin2019tasi} and evidence from both direct detection experiments and the Large Hadron Collider (LHC) at CERN are increasingly pointing towards dark matter being lighter than first thought, with masses less than that $\sim$ 1 GeV. 
\textbf{Helium}, being the second lightest element in nature, {\em is an excellent target material for detecting such light particles}. Previously helium was utilized as a target in the HERON neutrino detection experiment \cite{Lanou:1988iq,Bandler:1991ep,Bandler:1992zz,Adams:1996ge} 
and there are proposals for similar dedicated DM experiments in the future, e.g., HeRaLD \cite{Hertel_2019}.

To detect dark matter with superfluid helium, we need to understand how the dark matter interacts with helium in different mass ranges and how the reaction of the helium can be translated to a detector signal. Previous work in this field has primarily focused on masses above GeV \cite{Guo:2013dt,Ito_2013,bandler1992particle,maris2017dark} or below MeV \cite{Schutz_2016,Knapen_2017,Acanfora_2019,Caputo_2019,Caputo:2019xum,Caputo:2019ywq,baym2020searching,Caputo:2020sys}.  Heavier dark matter collisions are energetic enough to excite or ionize helium atoms, while lighter dark matter coherently scatters with the condensate, and the dominant process is one that produces only two phonons. Detection of individual phonons is currently beyond the reach of our techniques, but larger cascades of phonons and rotons will evaporate He atoms from the surface of the superfluid, enabling the HeRALD experiment to determine the total energy deposited in the quasiparticle modes \cite{Hertel_2019}. To better understand the helium response in the intermediate mass range between MeV and GeV,  we apply a ``cascade'' model where the initial dark matter collision excites a single energetic helium atom out of the condensate. This ``fast'' helium travels ballistically through the superfluid, elastically exciting new fast helium atoms out of the condensate, which then both continue to excite He atoms {\em and} radiate quasiparticles into the superfluid as they lose energy. The greater population of quasiparticles in these cascades should translate to improved experimental sensitivity, and several ideas for detection will be explored in future work \cite{MSXY}.
However, a proper estimate of experimental sensitivity in future experiments requires a thorough understanding of the physics of the resulting cascade, which is governed by the  interactions between the He atoms, the phonons and the rotons, which we develop here in the language of effective field theory (EFT).

\subsection*{Towards an effective theory for superfluid helium}

\begin{figure}[t]
  \centering
  \includegraphics[width=0.8\textwidth]{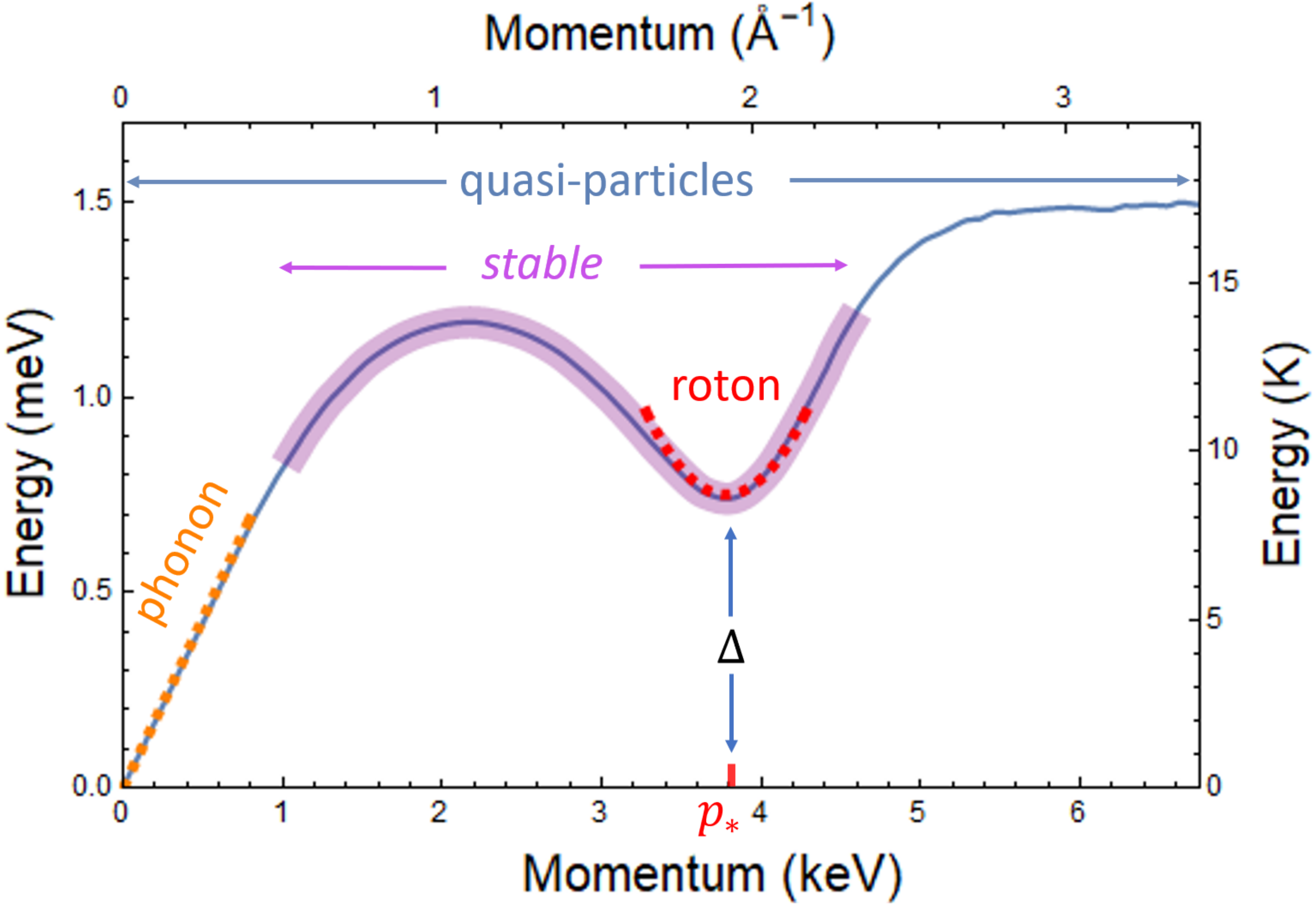}
    \caption{Dispersion curve for the  quasiparticles in  superfluid He measured in \cite{godfrin2021dis}. The orange dashed line is the linear phonon dispersion $E=c_s p$ with sound speed $c_s$. The red dashed line corresponds to the roton dispersion \cref{eq:rotonE}. The deep purple band is the region where quasiparticles are stable.}
\label{fig:dispersion}
\end{figure}

At temperatures below $T_\lambda =  \, 2.17 \, {\rm K}$, liquid $^4$He enters a new phase (He~$\RN{2}$) distinguished by its lack of viscosity (for reviews on helium superfluidity see \cite{Schmitt:2014eka,Vilchynskyy_2013}). This is a manifestation of Bose Einstein condensation \cite{BOGOYAVLENSKII1992151,Wyatt_1998}, deriving from the exchange symmetry of spinless $^4$He atoms.
In the superfluid phase, perturbations in the fluid density are quantized and may be described as ``quasiparticles'', particle-like excitations of the density field. The spectra and dynamics of these excitations are the subject of extensive investigation on both the experimental and theoretical fronts.  As shown in \cref{fig:dispersion}, the quasiparticles have a linear dispersion branch at low momenta, where they are known as \textbf{phonons}. In 1941, Landau constructed a phenomenological theory to understand the spectrum and dynamics of these quasiparticles \cite{landau1941theory,khalatnikov2018introduction}. In this phenomenological theory, 
phonons are the fields associated with the quantization of density and velocity of superfluid.
In addition Landau posited the existence of the roton to explain the anomalous specific heat of He~$\RN{2}$; later the \textbf{roton} excitations were understood as a higher momentum branch, the local minimum around $4 \, {\rm keV}$ in \cref{fig:dispersion}, of the same quasiparticle dispersion relation. (The whole branch in Fig.~\ref{fig:dispersion} is collectively referred to as \textbf{quasiparticles}.) Measurements of energy and momentum loss of slowly moving neutrons in a superfluid helium target determine the dynamic response of the superfluid \cite{henshaw1961modes}.  

There are several obstacles to the construction of a fundamental theory for superfluid excitations. The presence of the superfluid breaks Lorentz invariance, and strong couplings between helium atoms complicate the use of perturbation theory.  Nevertheless, there exists a separation of scales between the energies of phonons, rotons, and fast moving neutral probes (which could be neutrons, dark matter particles or He atoms), which inspires us to construct an effective field theory for all these degrees of freedom. Effective field theory is an appropriate theoretical tool to compute scattering amplitudes and making predictions. 
The strong couplings of helium and quasiparticles as the low energy degrees of freedom make the situation close to low energy QCD,
in which the quarks and gluons are strongly coupled and mesons can be described by a perturbative effective field theory \cite{weinberg_1995,Schwartz:2014sze}. 
Several new frameworks have been developed to study the dynamics of mesons, baryons, heavy quarks, etc.
These new developments can potentially shed insights on the physics of helium superfluid and motivate us to construct a novel theoretical framework for 
helium superfluid by using an effective field theory.
Previous work on the effective field theory of superfluid helium has focused on low energy quasiparticles lying deep in the linear dispersion regime, where the phonon arises as the Nambu-Goldstone boson of a U(1) symmetry broken by the helium Bose Einstein condensate \cite{Greiter:1989qb,son2002lowenergy,nicolis2011low,leutwyler1996phonons}. In such a scheme rotons are treated as background particles sourcing phonon fields, but their own dynamics are not treated \cite{Nicolis_2018}.

In this paper, we construct an effective field theory framework to describe the dynamics of helium superfluid under external probes. A systematic study of the direct detection signal in superfluid He due to a dark matter particle (focusing specifically on the challenging and unexplored mass range from 1 MeV to 1 GeV) will be presented in a future work \cite{MSXY}.
The degrees of freedom that we are considering here include: quasiparticles and helium atoms as constituents of the cascade, as well as high momentum helium atoms, neutrons and dark matter as probe particles.
The effective field theory for phonons as Goldstone modes has been developed in  \cite{Greiter:1989qb,son2002lowenergy,Nicolis_2018,nicolis2011low,leutwyler1996phonons}. 
We will follow the relativistic superfluid method given in \cite{son2002lowenergy} and generalize to non-relativistic helium superfluid.  
For rotons, we will use the power-counting method in effective field theories (see \cite{polchinski1992effective} for Fermi liquid)
to derive the effective Lagrangian. To further determine some of the coefficients for roton-phonon interactions, we will consider rotons 
as impurities in the superfluid. In analogy to the baryon couplings to pions, the interactions of helium atoms with phonons can be deduced by 
considering the $U(1)$ symmetry of non-relativistic helium atoms. To describe the radiation of quasiparticles by helium atoms,  we will treat the helium atoms as neutral probes and rely on the measurement of neutron momentum loss in the superfluid.

The organization of our paper is as follows. In \cref{sec:setup}, we discuss
the relevant degrees of freedom and processes we will consider
in the superfluid $^{4}$He. In \cref{sec:theory}, we construct the effective field theory
for quasiparticles and other external probes and discuss their application for the calculation of processes of phenomenological interest. To make contact with previous work, in \cref{sec:crosssecton} we reproduce the relevant results for decay rates and cross-sections and summarize them in 
\cref{sec:summarycs}.
Our conclusions and outlook for future studies are given in \cref{sec:conclusions}.

\section{Relevant degrees of freedom and processes}
\label{sec:setup}
In this section we introduce the relevant degrees of freedom and estimate the UV cutoff of the theory. We will consider quasiparticles (which include phonons, rotons, as well as other excitations lying along the same dispersion curve), helium atoms, and probe particles, such as dark matter and neutrons.

\subsection*{Degrees of freedom}

In the non-relativistic limit, helium atoms $\Phi$ obey the Lagrangian:
\begin{equation}
   {\cal L}_{He}=\Phi^\dagger i \partial_t \Phi 
    - \frac{1}{2 m_{\rm He}} \partial_i\Phi^\dagger
      \partial_i \Phi \ ,
   \label{eq:LHe0}
\end{equation}
which is invariant under the $U(1)$ transformation $\Phi \to {\rm e}^{ - i \alpha} \, \Phi$.
The associated Noether current is 
\begin{equation}
   J^{\mu} = \left( \Phi^\dagger \Phi \, , \, \frac{-i} { 2 m_{\rm He}} (\Phi^\dagger \partial_i\Phi -
   \Phi \partial_i \Phi^\dagger ) \right)  \ , 
   \label{eq:Jmu}
\end{equation}
with the zero component $J^0$ the number density of helium atoms. At temperature of about $2 \, {\rm K}$,
the helium undergoes a phase transition and a fraction of it forms a Bose Einstein condensate,
which breaks both this $U(1)$ symmetry and the boost symmetry of the system.
The quasiparticles are 
the Nambu-Goldstone modes associated with this broken $U(1)$ symmetry, as we will elaborate in the next section.

A distinctive dispersion relation characterizes quasiparticle excitations
in the superfluid $^{4}$He, as shown in \cref{fig:dispersion}.
``Phonons'' refers to the quasiparticles with momenta below $\sim 2\ {\rm keV}$. 
In this region the dispersion relation is linear in $p$ with a small cubic correction:
 \begin{equation}
    E_\text{phonon} \simeq c_s \, \left( p   \, - \, \frac{ \gamma } {  \Lambda^2  }   \, p^3 \right),
   \label{eq:phononE}
 \end{equation}
where $c_s = 240 \, {\rm m/s}$ is the speed of sound in the superfluid, $\Lambda$ is the UV scale of the superfluid helium, and $\gamma \sim {\cal O} (1) $  is a dimensionless parameter, such that the combination 
$\gamma/\Lambda^2 =0.27 {\buildrel _{\circ} \over {\mathrm{A}}}{}^2    $.
Near the local minimum at momentum $p_\ast = 3.84$ keV, the quasiparticle's energy is parametrized as 
\begin{equation}
E_{roton}\simeq\Delta+\frac{(p-p_{*})^{2}}{2m_{*}} 
   \label{eq:rotonE}
\end{equation}
and the corresponding excitations are called ``rotons'' (see the red dashed line in \cref{fig:dispersion}). Here $\Delta = 0.75$ meV is the gap energy and $m_\ast \simeq 0.16\, m_{\text{He}}$ is the effective roton mass parameter.
All parameters in eqs.~(\ref{eq:phononE}) and (\ref{eq:rotonE}) are thermodynamic quantities which depend on the superfluid's temperature and density --- 
the quoted values here are extrapolated to zero pressure and density $0.145\ {\rm g}/{\rm cm}^3$ \cite{landau1941theory}.

We can estimate the UV cutoff $\Lambda$ in several ways, all leading to consistent results. The inverse atomic spacing in the superfluid yields $\Lambda \sim {\rm keV}$. Inspection of the dispersion relation \cref{eq:phononE} shows an ${\cal O}(1)$ deviation from linearity at $2$ keV. Furthermore, the energy associated with the effective roton mass $c_s m_* \sim {\rm keV}$ also falls at the same scale.

Due to interactions between quasiparticle modes, the lifetimes of quasiparticles are sensitive to their momenta. In the central region of $1 \, {\rm keV} \, - \, 4.5 \, {\rm keV}$, the quasiparticles are stable,
as the dispersion relation forbids the decay kinematics. Softer phonons of $q <  1 \, {\rm keV}$, while unstable, may only decay to indistinguishable configurations of collinear phonons with the same total energy and momentum as the parent phonon.

We will consider other probe particles that may scatter off the superfluid, such as neutrons and dark matter particles. These probe particles are neutral both electrostatically and under the helium $U(1)$ symmetry, but their interactions with the superfluid may be derived from coupling to the helium number current. Recoiling helium atoms from the cascade may also act as probe particles.

\subsection*{Processes}

As we are motivated by dark matter direct detection, we consider here the initial process which triggers the cascade to be the scattering of a dark matter particle with mass in the range 1 MeV to 1 GeV off the superfluid. Because the de Broglie wavelength of the dark matter particle is shorter than the average atomic spacing of helium, we expect the initial dark matter collision to excite a single ``fast'' helium atom out of the condensate. The main focus of this work is to develop the theoretical framework needed to describe the physics of the subsequent helium cascade. From this point of view, we shall be agnostic to the specific dark matter model which was responsible for the initial collision, and shall parametrize its outcome simply with the momentum of the recoiling helium atom. A complete understanding of the helium cascade then requires description of  
$(1)$ helium-helium scattering, $(2)$ emission of quasiparticles by helium atoms, and $(3)$ quasiparticle decays and scattering.

Given the helium-helium potential, the helium-helium scattering phase shifts can be derived using perturbative quantum mechanical calculations at low momenta and semi-classical methods at high momenta (see ~\cref{sec:appendix}). 

In the next sections, we will employ the effective field formalism to study the following quasiparticle processes:
\begin{itemize}
\item  emission of quasiparticles by helium atoms,
\item unstable quasiparticle decays to lower momentum quasiparticles,
\item stable quasiparticles scattering off each other.
\end{itemize}

\input{theory}

\input{process}




\section{Conclusions and outlook}
\label{sec:conclusions}

\input{conclusions}

\acknowledgments

We are grateful to Wei Guo, Simon Knapen, Yoonseok Lee, Tongyan Lin, Dmitrii Maslov, Riccardo Penco, Tarek Saab and Edward Witten for useful discussions. 
This work was supported in part by the United States Department of Energy
under Grant No. DE-SC0010296.


\appendix

\input{appendix}

\bibliographystyle{JHEP}
\bibliography{ref}

\end{document}

%% file: theory.tex
\section{Effective field theories for superfluid}
\label{sec:theory}

In this section we study the relevant interactions among phonons, rotons, fast-moving helium atoms, and neutral probe particles, 
and use symmetry, power counting arguments and impurity method to construct effective Lagrangians describing their dynamics.

\subsection{Phonon as a Goldstone boson in the superfluid}

\label{sec:phonon}

Helium atoms are strongly coupled at or below ${\rm keV}$, while phonons are massless Goldstone modes which can be described by a perturbative effective field theory.
We construct the effective Lagrangian for phonon interactions with two methods: one is based on the symmetry of the quantum action \cite{son2002lowenergy}, while the other is a simple power-counting method (many of those results have previously appeared in refs.~\cite{landau1941theory,Greiter:1989qb,Nicolis_2018,Caputo:2020sys}).
We note that
the power-counting method cannot completely fix
the coefficients, thus it is only mentioned here for completeness.

The Goldstone modes emerge because the superfluid background breaks the $U(1)$ symmetry associated with the number of helium atoms.
This is expressed by first adding to the Lagrangian in \cref{eq:LHe0} a chemical potential $\mu$ coupled to the helium number density: 
\begin{align}
{\cal L} & ={\cal L}_{He}\,+\,\mu\,\Phi^{*}\Phi
 ~~\longrightarrow ~~ {\cal L}_{He}\,+\,A_{\nu}J^{\nu}\ ,
\end{align}
where we generalize $\mu\,\Phi^{*}\Phi$ to $A_{\nu}J^{\nu}$ with the background field values $A_{0}=\mu$ and $A_{i}=0$ and $J^\nu$ is defined as in \cref{eq:Jmu}. We promote this background field to a spurious gauge field and subject the broken symmetry to a ``spurion'' analysis. The Lagrangian augmented with the background field term then respects the symmetry
\begin{equation}
\Phi\to e^{-i\alpha(\boldsymbol{x},t)}\Phi,\quad
A_{\mu}\to A_{\mu}-\partial_{\mu}\alpha(\boldsymbol{x},t)\ .
\end{equation}
This gauge symmetry is also respected in the quantum effective action $\Gamma(\Phi,A_{\mu})$.
In the superfluid background we are examining, the modulus $\vert \Phi \vert$ will take on a vacuum expectation value and the remaining degree of freedom, the complex phase, is identified with the phonon field $\pi_{0}$ such that $\Phi=|\Phi|e^{-i\pi_{0}}$. Under this field definition, the spurious gauge symmetry of $\Phi$ is transformed into a shift symmetry in $\pi_0$:
\begin{equation}
\Gamma(\pi_{0},A_{\mu})=\Gamma(\pi_{0}+\alpha,A_{\mu}-\partial_{\mu}\alpha)\ .
\end{equation}
To ensure the invariance of the effective action, $\pi_0$ must only appear as part of the Galileian and (spurious) gauge invariant combination\footnote{An alternative relativistic derivation of the same phonon effective Lagrangian is given in \cite{Acanfora_2019}.}  
\begin{equation}X=\mu+\dot{\pi}_{0}-\frac{\partial_{i}\pi_{0}\partial_{i}\pi_{0}}{2m_{\text{He}}}.
\end{equation}
To connect this discussion of symmetries to the dynamics of phonon interactions we need an additional external ingredient: we notice the background effective Lagrangian is proportional to the background energy density, which itself is proportional to the thermodynamic pressure $P$ of the fluid. Therefore the Lagrangian of the superfluid system is just its pressure
\begin{equation}
{\cal L}_\text{eff}=P(X)=P\left(\mu+\dot{\pi}_{0}-\frac{\partial_{i}\pi_{0}\,\partial_{i}\pi_{0}}{2m_{\text{He}}}\right)\ ,\label{eq:chemexpan}
\end{equation}
and deviations from a free theory of phonons are determined by the superfluid equation of state $P(X)$.

We can now Taylor expand $P(X)$ about the equilibrium value $X=\mu$ to obtain the phonon interactions:
\begin{align}
{\cal L}_{ph} & =P(\mu)+P^{\prime}(\mu)
\left(\dot{\pi}_{0}-\frac{\partial_{i}\pi_{0}\,\partial_{i}\pi_{0}}{2m_{\text{He}}}\right)+\frac{1}{2}P^{\prime\prime}(\mu)
\left(\dot{\pi}_{0}-\frac{\partial_{i}\pi_{0}\,\partial_{i}\pi_{0}}{2m_{\text{He}}}\right)^{2}\nonumber \\
 & +\frac{1}{6}P^{\prime\prime\prime}(\mu)\left(\dot{\pi}_{0}-\frac{\partial_{i}\pi_{0}\,\partial_{i}\pi_{0}}{2m_{\text{He}}}\right)^{3}+\frac{1}{24}P^{\prime\prime\prime\prime}(\mu)
\left(\dot{\pi}_{0}-\frac{\partial_{i}\pi_{0}\,\partial_{i}\pi_{0}}{2m_{\text{He}}}\right)^{4}+\cdots\ .\label{eq:Leff_phonon}
\end{align}
The coefficients $P^{\prime}$, $P^{\prime\prime}$, $P^{\prime\prime\prime}$ and $P^{\prime\prime\prime\prime}$
are determined by measurements of the superfluid equation of state:
\begin{equation}
    \begin{split}
        P^{\prime} & \equiv\frac{{\rm d}P}{{\rm d}\mu}=\frac{\rho}{m_{\text{He}}}\,, \\
        P^{\prime\prime}& \equiv\frac{{\rm d}^{2}P}{{\rm d}\mu^{2}}=\frac{\rho}{\mu\,m_{\text{He}}}
        =\frac{\rho}{m_{\text{He}}^2c_s^2} \\
        P^{\prime\prime\prime} & \equiv\frac{{\rm d}^{3}P}{{\rm d}\mu^{3}}=\frac{{\rm d}P^{\prime}}{{\rm d}\mu}\,\frac{{\rm d}P^{\prime\prime}}{{\rm d}P^{\prime}}=\frac{\rho}{m_{\text{He}}^{3}\,c_{s}^{4}}\,\left(1-2u
        \right)\,, \\
        P^{\prime\prime\prime\prime} & \equiv\frac{{\rm d}^{4}P}{{\rm d}\mu^{4}}=\frac{\rho}{\mu}\,\frac{{\rm d}P^{\prime\prime\prime}}{{\rm d}\rho}=\frac{\rho}{m_{\text{He}}^{4}\,c_{s}^{6}}\,\left(1-8u+10u^{2}-2\omega\right)\ ,
    \end{split}
    \qquad
    \begin{split}
        \quad & \\
        c_{s}^{2}&=\frac{m_{\text{He}}\,P^{\prime}}{P^{\prime\prime}}=\frac{\mu}{m_{\text{He}}}\ , \\
        u &=\frac{\rho}{c_{s}}\frac{{\rm d}\,c_{s}}{{\rm d}\rho}=2.84\,, \\
        \omega &=\frac{\rho^{2}}{c_{s}}\frac{{\rm d^{2}}\,c_{s}}{{\rm d}\rho^{2}}=8.26\,,
    \end{split}
    \label{eq:Pprimes}
\end{equation}
where the numerical values are given in \cite{abraham1970velocity}.
Thus the phonon interactions are fixed by the superfluid sound speed $c_{s}$, energy density $\rho$, and helium mass $m_{\text{He}}$. We perform a field rescaling such that the Lagrangian is canonically normalized: 
\begin{equation}
\pi=\frac{\sqrt{\rho}}{m_{\text{He}} ~c_{s}}\,\pi_{0}\,.\label{eq:pi0topi}
\end{equation}

In the effective field theory language, non-renormalizable Lagrangian coefficients are expressed
in inverse powers of the UV cutoff
\begin{equation}
\Lambda\equiv(\rho\,c_{s})^{1/4}\,,\label{eq:lamdef}
\end{equation}
at which we expect our perturbative calculations to break down. This value of the cutoff is consistent with the approximate scalings
 $n\sim\Lambda^{3}$ and $c_{s}\sim\Lambda/m_{\text{He}} $. 
With these simplifying definitions, after substituting
eqs.~(\ref{eq:Pprimes}-\ref{eq:lamdef})
into  \cref{eq:Leff_phonon}, our Lagrangian becomes, up to quartic interactions 
\begin{align}
{\cal L}_{{\rm ph}} & =\frac{1}{2}\left(\dot{\pi}^{2}-c_{s}^{2}\,\partial_{i}\pi\partial_{i}\pi\right)-\frac{c_{s}^{3/2}}{2\Lambda^{2}}\,\dot{\pi}\,\partial_{i}\pi\,\partial_{i}\pi
+\frac{g_{3}\,c_{s}^{-1/2}}{6\,\Lambda^{2}}\,\dot{\pi}^{3}\nonumber \\
 & +\frac{c_{s}^{3}}{8\,\Lambda^{4}}\left(\partial_{i}\pi\,\partial_{i}\pi\right)^{2}
-\frac{g_{3}\,c_{s}}{4\,\Lambda^{4}}\,\dot{\pi}^{2}\,\partial_{i}\pi\,\partial_{i}\pi+\frac{g_{4}\,c_{s}^{-1}}{24\,\Lambda^{4}}\,\dot{\pi}^{4}+\cdots ,\label{eq:Lphononinteraction}
\end{align}
where the couplings $g_3$ and $g_4$ are proportional to $P^{\prime\prime\prime}$
and $P^{\prime\prime\prime\prime}$, respectively: 
\begin{equation}
g_{3}=1-2u\ ,\quad
g_{4}=1-8u+10u^{2}-2\omega\ .
\end{equation}

The powers of $c_{s}$ and $\Lambda$ appearing in \cref{eq:Lphononinteraction} can be alternatively motivated by the following power counting method. 
First, using mass dimensions, we have
\begin{equation}
\quad[p]=1\,,
\quad[\partial_{i}]=1\,,
\quad[t]=-1\,,\quad[\partial_{t}]=1\,,\quad[x]=-1\,,\quad[\pi]=1\ ,\label{eq:phononcounting}
\end{equation}
which can be used to obtain the power of the cut-off scale $\Lambda$ in each operator. The powers of $c_{s}$ can then be obtained from the SI units. This leaves dimensionless coefficients of order 1 which are fixed by symmetries or observations.

\subsection{Roton interactions}
\label{sec:roton}

Next we construct an effective theory of roton interactions. In most conventional EFTs amplitudes are organized as a power series in $p/\Lambda$.
However, the roton momentum is about $p_* = 3.84\,{\rm keV}$, which is larger than the UV cutoff we introduced in the phonon EFT in the previous section. To proceed, we borrow philosophically from theories like the EFT of the Fermi liquid \cite{polchinski1992effective} or of heavy quarks, where momenta are defined as\footnote{From here on, boldface notation will indicate a 3-vector $\mathbf{p}$, with its magnitude denoted by $p$ or $|\mathbf{p}|$.}  ${\bf p}={\bf k}+\boldsymbol{\ell}$, namely, in terms of small deviations $\boldsymbol{\ell}$ from a large reference momentum ${\bf k}$ (the Fermi surface or the heavy quark momentum). Amplitudes are then expressed as a power series in $\ell/\Lambda$ rather than $p/\Lambda$. In the same spirit, we define a roton theory in terms of small deviations $\boldsymbol{\ell}$ from the characteristic roton momentum $\boldsymbol{p_*}$ and determine the interaction structure from power counting techniques. To fix the interaction coefficients, in \cref{sec:impurity} we match to macroscopic fluid dynamics descriptions of superfluid helium \cite{landau1941theory,khalatnikov2018introduction}.

\subsubsection{Roton self-interaction}

As illustrated in \cref{fig:dispersion}, the roton dispersion relation has a local minimum of $\Delta$ at the momentum surface $\vert {\bf p} \vert = p_*$. As such we can write all roton momenta in terms of small deviations $\boldsymbol{\ell}$ about the characteristic roton momentum ${\bf p}_{*}$, where $\boldsymbol{\ell}$ is parallel to ${\bf p}_{*}$:
\begin{equation}
{\bf p}={\bf p}_{*}+\boldsymbol{\ell}=(p_{*}+\ell)\,\hat{\mathbf{p}}\,,
\end{equation}
where $\hat{\mathbf{p}}=\mathbf{p}/p$ is a unit vector in the direction of ${\bf p}$.

We write the roton as a scalar field $\varphi_{r}$ and apply a phase shift to factor out the large plane wave component defined by $p_*$ and $\Delta$. The remaining factor $\tilde{\varphi}$ represents quantum fluctuations about this classical trajectory. In the absence of interactions this phase-shifted field will obey the action
\begin{equation}
S_{\text{free}}[\varphi_{r}]=\int dt\frac{d^{3}p}{(2\pi)^3}\left[\tilde{\varphi}_{r}^{\ast}({\bf p})\,i\partial_{t}\,\tilde{\varphi}_{r}({\bf p})-\frac{\ell^{2}}{2m_{\ast}}\,\tilde{\varphi}_{r}^{\ast}({\bf p})\,\tilde{\varphi}_{r}({\bf p})\right],\label{eq:rotonfree}
\end{equation}
where $\tilde{\varphi}_{r}({\bf p})$
is the Fourier transformed $\tilde{\varphi}_{r}({\bf r})$. We use this free action to determine the dimension of the roton field: to maintain invariance of the action under scaling the residual roton momentum $\boldsymbol{\ell}\to s\boldsymbol{\ell}$, other ingredients of the action must scale appropriately. Because we scale only the component of the momentum normal to the $d^2 p$ surface element, while keeping $p_*$ fixed, the infinitesimal phase space element $d^3p$ carries scaling dimension 1 rather than 3. From this we deduce the scaling behavior of the other variables in the action:
\begin{equation}
[\ell]=1\,,\quad[p_{*}]=0\,,\quad[{\rm d}^{3}p]=1\,,\quad[\tilde{\varphi}_{r}]=-\frac{1}{2}\,,\quad[{\rm d}t]=-2\,,\quad[\partial_{t}]=2 \ .\label{eq:rotonpower}
\end{equation}

Now it is straightforward to write down the leading roton interaction operator consistent with these scaling dimensions
\begin{align}
S_{\text{int}}[\varphi_{r}]= & \frac{1}{(2\pi)^9} \int{\rm d}t({\rm d}^{2}{\bf p}_{\ast})^{4}{\rm d}\boldsymbol{l}_{1}{\rm d}\boldsymbol{l}_{2}{\rm d}\boldsymbol{l}_{3}{\rm d}\boldsymbol{l}_{4}\,\,\delta^{3}(\boldsymbol{p}_{1}+\boldsymbol{p}_{2}-\boldsymbol{p_{3}-}\boldsymbol{p}_{4})\nonumber \\
 & \times\frac{\lambda_{r}({\bf p}_{1},{\bf p}_{2},{\bf p}_{3},{\bf p}_{4})}{m_{*}\,p_{*}}\,\tilde{\varphi}_{r}^{\ast}({\bf p}_{1})\,\tilde{\varphi}_{r}({\bf p}_{2})\,\tilde{\varphi}_{r}^{\ast}({\bf p}_{3})\,\tilde{\varphi}_{r}({\bf p}_{4}),
\label{eq:rotoninteractions}
\end{align}
where $\boldsymbol{p}_{i}$ and $\boldsymbol{l}_{i}$ are the momenta associated with the $i$th roton. The coupling $\lambda_r$ is a scalar and is generally a function of the roton momenta, but here we assume it is a constant and use the value $\lambda_{r}=0.93$ \cite{1965511}. The prefactor $1/(m_* p_*)$ is obtained by dimensional analysis. Note that this operator is marginal under the power counting scheme outlined in \cref{eq:rotonpower}, while in the nonrelativistic limit of traditional $\varphi^4$ theory this operator is irrelevant.  

\subsubsection{Roton-phonon interaction}

We use power counting to determine the form of the roton-phonon interactions, restricted to 2 external rotons and 1 or 2 external phonons. In contrast with the previous section it is more convenient to write this Lagrangian in position space, but we have verified that power counting in momentum space yields the same results. The scaling dimensions are
\begin{equation}
[t]=-1\,,\quad[x]=-1\,,\quad[{\varphi}_{r}]=\frac{3}{2}\,,\quad[\pi]=1\,.
\label{eq:rotonphononpower}
\end{equation}
Note the scaling dimension of $t$ is the same as in the phonon interactions of \cref{eq:phononcounting} but differs from the roton scheme in \cref{eq:rotonpower}. This apparent discrepancy is resolved by recalling that $m_*$ and $p_*$ do not scale. Taking the approximation that $m_* \sim \Lambda/c_s$ and $p_* \sim \Lambda$, the effective couplings are expressed as\footnote{These results also appear (in a somewhat different form) in \cite{2017PhRvL.119z0402C,Nicolis_2018}.} 
\begin{align}
\mathcal{L}_{\text{ph-r}} & =\frac{y_{3}\,c_{s}^{1/2}}{\Lambda}\,{\varphi}_{r}^{*}\,{\varphi}_{r}\,\dot{\pi}+\frac{y_{4}\,}{2\,\Lambda^{3}}\,{\varphi}_{r}^{*}\,{\varphi}_{r}\,\dot{\pi}^{2}+\frac{y_{4}^{\prime}\ c_{s}^{2}}{2\,\Lambda^{3}}\,{\varphi}_{r}^{*}\,{\varphi}_{r}\,\partial_{i}\pi\,\partial_{i}\pi\nonumber \\
 & +\left(\frac{y_{3}^{\prime}\,c_{s}^{3/2}}{\Lambda^{2}}\,{\varphi}_{r}^{*}\partial_{i}{\varphi}_{r}\,\partial_{i}{\pi}+\frac{y_{4}^{\prime\prime}\,c_{s}}{\Lambda^{4}}\,{\varphi}_{r}^{*}\partial_{i}{\varphi}_{r}\,\partial_{i}{\pi}\,\dot{\pi}+\,h.c.\right)\nonumber \\
 & +\frac{y_{4}^{\prime\prime\prime}\,c_{s}^{2}}{2\,\Lambda^{5}}\,\partial_{i}{\varphi}_{r}^{*}\,\partial_{i}\pi\partial_{j}{\varphi}_{r}\,\partial_{j}{\pi}\ .\label{eq:rotonphononinteraction}
\end{align}
Here we obtain the power of $c_{s}$ from the SI units as was done in the case of the phonon self-interactions in \cref{eq:Lphononinteraction}.

The coupling constants in this Lagrangian are fixed using a fluid dynamics matching technique we will demonstrate in the next section. In terms of the physical parameters $\Delta$, $\rho$, $p_*$, $m_*$, and $c_s$ they are
\begin{equation}
    \begin{split}
        y_{3} & =-y_{4}'=c_{s}^{-\frac{5}{4}}\rho^{\frac{3}{4}}\frac{d\Delta}{d\rho}, \\
        y_{3}^{\prime} & =\frac{i}{2}, \\
        y_4^{\prime\prime\prime} & = 0, 
    \end{split} 
    \qquad
    \begin{split}
        y_{4} &=c_{s}^{-\frac{5}{4}}\rho^{\frac{7}{4}}\left[\frac{d^{2}\Delta}{d\rho^{2}}+\frac{1}{m_{\ast}}\left(\frac{dp_{\ast}}{d\rho}\right)^{2}\right], \\
        y_{4}^{\prime\prime}&=\frac{i}{2}\frac{\rho}{p_{\ast}}\frac{d p_{\ast}}{d\rho}, \\
    \quad
    \end{split}
    \label{eq:rotonphononinteractionC}
\end{equation}
where the derivatives can be estimated experimentally \cite{peshkovJETP}: $d^{2}\Delta/d\rho^{2}\simeq2\Delta/\rho^{2}$
and $dp_{\ast}/d\rho\simeq p_{\ast}/\rho$. The derivative $d\Delta/d\rho$ is negligible due to anomalously small thermal expansion coefficient of $^4\text{He}$ \cite{landau1949theory}, thus we drop this term in the matrix element calculation.

\subsubsection{Impurity interaction with fluid}
\label{sec:impurity}

The interactions between an impurity and phonons can be determined by a classical fluid dynamics argument \cite{landau1949theory,landau1987statistical}.
We fix some of the couplings of rotons and phonons in \cref{eq:rotonphononinteraction} by treating the roton as an impurity, while
the impurity method is more general, and the impurity may be a neutron, a dark matter particle, or other quasiparticles. 
We will first derive a general form of the interaction potential using the impurity method, then apply it to several examples.

\begin{figure}[tbp]
  \centering
  \includegraphics[width=0.8\textwidth]{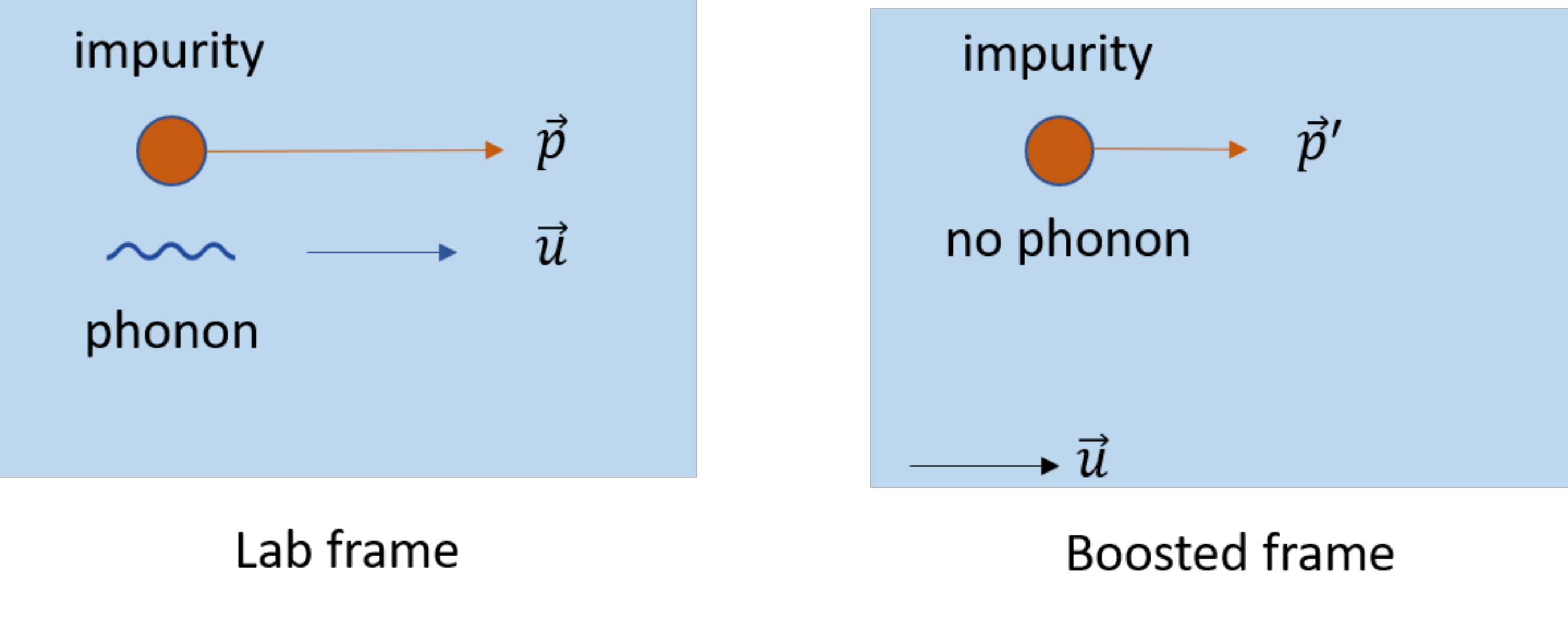}
    \caption{Impurity and phonon in a superfluid. The left system is observed in the lab frame, while the right system is in a boosted frame with vanishing phonon velocity.}
\label{fig:superfluid}    
\end{figure}

Imagine a superfluid bulk of mass $M$, carrying an impurity of vacuum mass $m$ with momentum $\boldsymbol{p}$ and a phonon of momentum $\mathbf{k}=k\,\hat{\mathbf{k}}$ and velocity $\boldsymbol{u}=c_s \boldsymbol{\hat{k}}$, as illustrated in the left side of fig. \ref{fig:superfluid}. The energy $E_L$ of this system in the lab frame is given by
\begin{equation}
    E_{L}=\epsilon(\boldsymbol{p},X_{0})
    + E_\text{ph}
    +V_{\text{ph-im}}, 
\label{eq:EL}
\end{equation}
where 
$$E_\text{ph}= c_s k=\boldsymbol{u}\cdot\boldsymbol{k}$$ is the phonon energy, and we have manually separated the free energy of the impurity $\epsilon(\boldsymbol{p},X_{0}) $ (which is the impurity's dispersion relation in a static fluid with $X_0$ the chemical potential) 
from the interaction potential $V_{\text{ph-im}}$
between the impurity and the phonon. 
Now boost the system by $\boldsymbol{u}$ so that the impurity carries a new momentum $\boldsymbol{p}^{\prime}=\boldsymbol{p}-m\boldsymbol{u}$ and the phonon is absent. The fluid bulk now contributes its free energy, and the total energy $E_B$ of the system in this boosted frame is 
\begin{equation}
   E_{B}=\epsilon(\boldsymbol{p}^{\prime},X)+Mu^{2}/2. 
\label{eq:EB}
\end{equation}
Under this Galilean transformation, the energy in the boosted frame is related to the energy in the lab frame by 
\begin{equation}
E_{B}=E_{L}-\left(\boldsymbol{p}+\boldsymbol{k}\right)\cdot\boldsymbol{u}+\frac{1}{2}\left(m+M\right)u^{2},\label{eq:dispersiontransformation}
\end{equation}
and thus the interaction potential can be determined: 
\begin{equation}
V_{\text{ph-im}}=\epsilon(\boldsymbol{p}-m\boldsymbol{u},X)-\epsilon(\boldsymbol{p},X_{0})+\boldsymbol{p}\cdot\boldsymbol{u}-\frac{1}{2}mu^{2}.\label{eq:IntPhIm}
\end{equation}

As an application of the above formalism, let us consider a few examples. The phenomenologically interesting case of a hard quasiparticle is postponed until the next \cref{sec:quasi-phonon}. Here we conclude the subsection with the three simpler examples of a completely non-interacting particle, a dark matter particle, or a roton. 

In the case where a classical particle simply does not interact with the superfluid, its dispersion relation $\epsilon \equiv p^2/2m$ is independent of $X$, and its interaction vanishes: 
\begin{equation*}
   V
= \epsilon(\boldsymbol{p}-m\boldsymbol{u})-\epsilon(\boldsymbol{p})+\boldsymbol{p}\cdot\boldsymbol{u}-\frac{1}{2}mu^{2} = 
    \frac{(\boldsymbol{p}-m\boldsymbol{u})^2}{2m}-
    \frac{\boldsymbol{p}^2}{2m}+\boldsymbol{p}\cdot\boldsymbol{u}-\frac{1}{2}mu^{2} = 0.
\end{equation*}

In the dark matter case, as a result of its interaction with the superfluid, the particle will acquire an effective mass $m_\text{eff}(X)$, which is a function of the chemical potential \cite{Acanfora_2019}. Under a perturbation in $X$ from its equilibrium value $X_0$, the dark matter Lagrangian acquires the mass correction $\left[m_{\text{eff}}^2(X)-m_{\text{eff}}^2(X_0)\right]|\chi|^{2}$, where $\chi$ is the dark matter field. In the nonrelativistic limit its dispersion relation changes according to 
\begin{equation}
    \epsilon(p,X)=\frac{p^2}{2m_{\text{eff}}(X_0)}+(\frac{1}{2}m_{\text{eff}}(X_0)-\frac{m_{\text{eff}}^2(X)}{2m_{\text{eff}}(X_0)}),
\end{equation}
and the potential is then
\begin{equation}
V_{\text{ph-DM}}=(\frac{1}{2}m_{\text{eff}}(X_0)-\frac{m_{\text{eff}}^2(X)}{2m_{\text{eff}}(X_0)})+\boldsymbol{p}\cdot\boldsymbol{u}(1-\frac{m}{m_{\text{eff}}(X_0)})-\frac{1}{2}(m-\frac{m^{2}}{m_{\text{eff}}(X_0)})u^{2}.
\label{eq:DMphEE0}
\end{equation}

The first term originates from the effective mass correction and thus the dark matter scalar coupling in the relativistic limit; note that the other terms are suppressed by factors of the dark matter or phonon velocities relative to the effective mass term.

Rotons have no rest mass and their dispersion relation is known to a good degree of numerical precision (see \cref{fig:dispersion}). The interaction potential with these conditions is 
\begin{equation}
V_{\text{ph-r}}=\epsilon(\boldsymbol{p},X)-\epsilon(\boldsymbol{p},X_{0})+\boldsymbol{p}\cdot\boldsymbol{u},
\label{eq:V_ph_ro}
\end{equation}

In order to specialize to the interactions with rotons we take the impurity momentum to be $p_*$. Under quantization the velocity operator is related to the spatial derivative of the phonon field
\begin{equation}
    \boldsymbol{u} \rightarrow c_s\boldsymbol{\nabla}\pi/\sqrt{\rho}
\end{equation}
and the deviation from equilibrium $X - X_0$ can be expanded in terms of the phonon field as in \cref{sec:phonon}. The roton parameters $p_*$ and $\Delta$ are themselves sensitive to the chemical potential, which introduces additional roton-phonon interactions that cannot be neglected. The end result was quoted in \cref{eq:rotonphononinteractionC}.
Note that this interaction potential can also be applied to other quasiparticles with their appropriate dispersion relations. 

We emphasize the utility of this formalism for fixing the coefficients of operators in the superfluid effective field theory. Power counting only allows us to determine the derivative structure of operators, but by matching to the nonrelativistic dynamics of impurities in the fluid we are able to connect the Lagrangian coefficients to experimental observables. Furthermore, this formalism allows us to argue simply why certain higher derivative interactions do not appear. For instance, we find $y_4^{\prime \prime \prime} = 0$ in \cref{eq:rotonphononinteraction,eq:rotonphononinteractionC} because the interaction potential contains no coupling of the form $(\boldsymbol{p}\cdot\boldsymbol{u})^2$.

\subsection{Quasiparticle and phonon interactions}
\label{sec:quasi-phonon}

Now we consider the interactions of quasiparticles with momenta beyond the roton range $q \gtrsim 5.5 ~\text{keV}$. Because these quasiparticles are unstable, a key question will be to determine their decay width. We follow the impurity approach introduced in \cref{sec:roton} to write the effective Lagrangian for these energetic quasiparticles interacting with phonons.

At momenta above $5.5 ~\text{keV}$ the quasiparticle dispersion relation approaches the approximate asymptotic value $2 \Delta$, which is roughly independent of momentum,
but still a function of the chemical potential: $\epsilon(\mathbf{p},X)\simeq\epsilon(X)$. As was the case with the roton field, we assign the quasiparticle a complex field $\varphi$ and factor out the plane wave component associated with this energy. The quasiparticle kinetic term fixes its scaling dimension to $[\varphi] = 3/2$, and power counting gives the same form of the interaction Lagrangian as the roton interactions in \cref{eq:rotonphononinteraction}, up to the values of $\mathcal{O}(1)$ coupling constants: 
\begin{equation}
\begin{split}
\mathcal{L}_{\text{ph-q}} & =\frac{z_{3}\,c_{s}^{1/2}}{\Lambda}\,{\varphi}^{*}\,{\varphi}\,\dot{\pi}+\frac{z_{4}\,}{2\,\Lambda^{3}}\,{\varphi}^{*}\,{\varphi}\,\dot{\pi}^{2}
\\ & 
+\frac{z_{4}^{\prime}\ c_{s}^{2}}{2\,\Lambda^{3}}\,{\varphi}^{*}\,{\varphi}\,\partial_{i}\pi\,\partial_{i}\pi+\left(\frac{z_{3}^{\prime}\,c_{s}^{3/2}}{\Lambda^{2}}\,{\varphi}^{*}\partial_{i}{\varphi}\,\partial_{i}{\pi}+h.c.\right).
\end{split}
\end{equation}

The impurity formalism fixes the couplings in terms of the thermodynamic dependence of the dispersion: 
\begin{align}
    z_{3}=-z_{4}'=c_{s}^{-\frac{5}{4}}\rho^{\frac{3}{4}}\frac{d\varepsilon}{d\rho} \,  ,
    \qquad
    z_{3}^{\prime}=\frac{i}{2} \, ,
    \qquad
    z_{4}=c_{s}^{-\frac{5}{4}}\rho^{\frac{7}{4}}\frac{d^{2}\varepsilon}{d\rho^{2}}\, ,
    \label{eq:z3}
\end{align}
while higher derivative terms analogous to the $y_4^{\prime \prime}$ and $y_4^{\prime \prime \prime}$ terms appearing in \cref{eq:rotonphononinteraction} vanish identically.

\subsection{Neutral particle emitting quasiparticles}
\label{sec:neutral2quasi}

Collective excitations of superfluid helium are probed by scattering neutral test particles (like neutrons) from the superfluid, and measuring their momentum and energy transfer. Our interest in these types of processes stems from the need to describe the helium response to sub-GeV dark matter scattering, which includes the emission of quasiparticles from similar neutral probes, namely, from dark matter particles themselves and from fast helium atoms generated from hard dark matter collisions. In this section we construct a unifying effective field theory framework for describing the emission of quasiparticles from all these neutral probes.

Previous experimental work has employed neutron scattering to measure the helium Dynamical Structure Function (DSF), while phenomenological work on the detection of sub-MeV dark matter with superfluid helium has treated the emission of phonons within an effective field theory framework 
\cite{Schutz_2016,Knapen_2017,Acanfora_2019,Caputo_2019,Caputo:2019xum,Caputo:2019ywq,baym2020searching,Caputo:2020sys}. Here we extend previous EFT work by observing that the DSF is related simply to the matrix element of the current 
$$S (\textbf{k}, \omega) \propto | \langle \varphi_\beta | J^0 | \Omega \rangle|^2, $$ 
which allows us to understand the emission of quasiparticles of all momenta.

In this section, we will first discuss the perturbative emission of quasiparticles by DM within the EFT framework and the associated matrix element $\langle \varphi_\beta | J^0 | \Omega \rangle$. Then we will treat neutron scattering in the same framework, connecting the effective interactions and the matrix element to the superfluid DSF. Finally, we shall discuss helium--quasiparticle interactions.

\subsubsection{DM emitting quasiparticles}

Consider the perturbative excitation of the superfluid by a DM particle. The interaction Lagrangian is well approximated by a coupling of the densities. This can be understood in terms of the impurity argument advanced in \cref{sec:impurity} and neglecting the terms suppressed by the quasiparticle velocity. Alternatively, we may recognize that very weak cross sections of the type that we anticipate from DM-initiated processes are well approximated by contact interactions with the helium atoms. The four-field local Lagrangian is then

\begin{equation}
\mathcal{L}_{\text{DM-qp}}=\frac{2\pi\lambda}{m_{\text{\ensuremath{\sigma}}}^{2}}\Phi_{\text{DM}}^{\ast}\Phi_{\text{DM}}\Phi_{\text{He}}^{\ast}\Phi_{\text{He}}=\frac{2\pi\lambda}{m_{\text{\ensuremath{\sigma}}}^{2}}j_{\text{DM}}^{0}J^{0},
\end{equation}
where $\Phi_{\text{DM}}$ and $\Phi_{\text{He}}$ are respectively the non-relativistic fields
of DM and helium;  $\lambda$ is a coupling constant, and $m_{\sigma}$
denotes a hypothetical mediator mass scale. Recall from \cref{eq:Jmu} that the number density is equal to the zero component of the probability current $J^{0}$, where $J^{\nu}$ is the Noether current of the superfluid $U(1)$ symmetry. At low momenta the timelike component of the current is approximately equal to 
$J^{0} \simeq \sqrt{\rho}\dot{\pi}/(m_{\rm He}c_{s})$,
where $\pi$ is the phonon field.

Let us write the external DM states as plane waves $|\boldsymbol{q}_{i}\rangle$
and $|\boldsymbol{q}_{f}\rangle$, the ground state of the superfluid as $|\Omega\rangle$ and the excited state as $|\varphi_{\beta}(\boldsymbol{k},\omega)\rangle$. Here $\boldsymbol{k}$ and $\omega$ denote the momentum and energy
transfer, respectively, and $\beta$ labels all possible single/multiple excitation configurations
carrying that energy and momentum. The matrix element of DM scattering is then 
\begin{align}
\mathcal{M}_{\beta} & =\langle\varphi_{\beta}(\boldsymbol{k},\omega)|\langle\boldsymbol{q}_{f}|\mathcal{L}_{\text{DM-He}}|\boldsymbol{q}_{i}\rangle|\Omega\rangle=\frac{2\pi\lambda}{m_{\text{\ensuremath{\sigma}}}^{2}}\langle\varphi_{\beta}(\boldsymbol{k},\omega)|J^{0}|\Omega\rangle.\label{eq:mbeta}
\end{align}
Note the matrix element factorizes into a dark matter contribution and the superfluid response to external density perturbations \cite{Schutz_2016,Knapen_2017}. While the dark matter has a definite momentum and energy, there are multiple possible superfluid states carrying the same energy and momentum. We must consider the contributions from all these configurations, which is schematically accomplished by summing over $\beta$. The quasiparticle emission rate from DM in the superfluid is then given by
\begin{align}
\Gamma_{\text{DM}} & =2\pi\left(\frac{2\pi\lambda}{m_{\text{\ensuremath{\sigma}}}^{2}}\right)^{2}\sum_\beta \int\frac{d^{3}k}{(2\pi)^{3}}\frac{|\langle\varphi_{\beta}(\boldsymbol{k},\omega)|J^{0}|\Omega\rangle|^{2}}{|\langle\varphi_{\beta}(\boldsymbol{k},\omega)|\varphi_{\beta}(\boldsymbol{k},\omega)\rangle|^{2}}\delta(E_{i}-E_{f}-\omega)\nonumber \\
 & =2\pi \frac{\rho}{m_\text{He}}\left(\frac{2\pi\lambda}{m_{\text{\ensuremath{\sigma}}}^{2}}\right)^{2}\int\frac{d^{3}k}{(2\pi)^{3}}S(\boldsymbol{k},\omega),
\end{align}
where the denominator in the integrand is the normalization factor
for quasiparticle states. The integrand is measured directly in neutron scattering experiments and known as the Dynamic Structure Function (DSF) $S(\boldsymbol{k},\omega)$, which is discussed further below.

\subsubsection{Neutron emitting quasiparticles}

\begin{figure}[htb]
  \centering
  \includegraphics[width=0.7\textwidth]{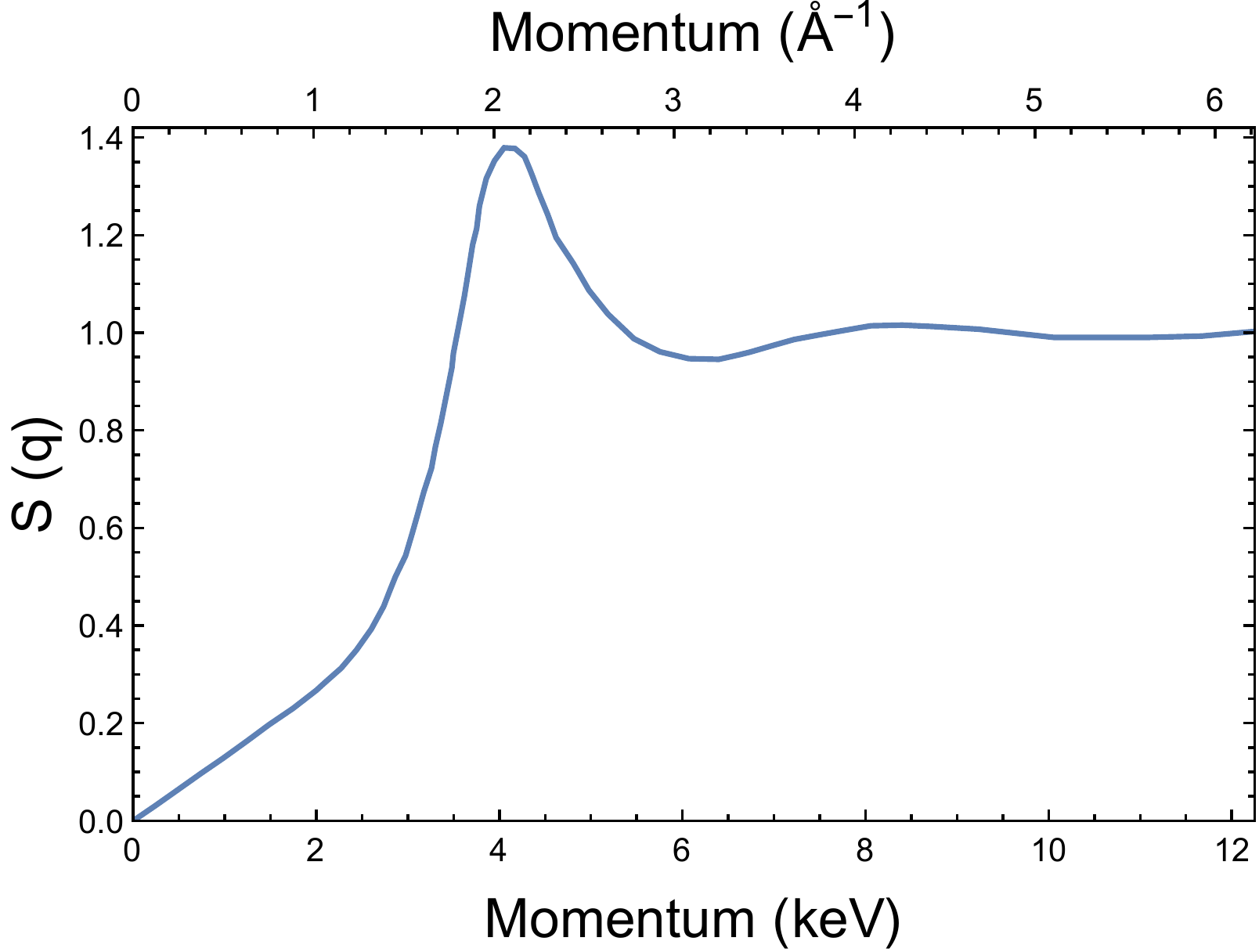}
    \caption{Interpolation of the data for the static structure function S(q) (at T = 1 K) from
neutron scattering experiment \cite{PhysRevB.21.3638}.}
\label{fig:ssf}    
\end{figure}

Let us briefly review the derivation of the DSF in the formalism of many body theory and link our EFT of quasiparticle currents to the matrix element of quasiparticle emission. Similar to the case of DM direct detection, the neutron kinetic energy is much larger than its interaction potential with helium, which can be approximated by a contact interaction $\propto \delta(\boldsymbol{x})$. Following the treatment of many body physics, we quantize the quasiparticle as fluctuations in the product Hilbert space of all the helium atoms in the superfluid. Let $\boldsymbol{R}_n$ denote the neutron position, $\boldsymbol{r}_a$ the position of the $a$-th helium atom, $b_n$ the neutron-helium scattering length, and $m_n$ the neutron mass. The potential between the neutron and the superfluid is expressed as a sum over all the helium atoms:
\begin{equation}
\mathcal{V}=\sum_{a}V(\boldsymbol{R}_{n}-\boldsymbol{r}_{a})=\frac{2\pi b_{n}}{m_{n}}\sum_{a}\delta^{3}(\boldsymbol{R}_{n}-\boldsymbol{r}_{a}).
\end{equation}
We are interested in the matrix element of this potential in the neutron-helium scattering process,
\begin{align}
M_{\beta} & =\langle\varphi_{\beta}(\boldsymbol{k},\omega)|\langle\boldsymbol{q}_{f}|\mathcal{V}|\boldsymbol{q}_{i}\rangle|\Omega\rangle=\frac{2\pi b_{n}}{m_{n}}\langle\varphi_{\beta}(\boldsymbol{k},\omega)|n_{\boldsymbol{k}}|\Omega\rangle,\label{eq:mbeta-1}
\end{align}
where $n_{\boldsymbol{k}}=\sum_{a}e^{i\boldsymbol{k}\cdot\boldsymbol{r}_{a}}$
is the Fourier transformed number density operator in the superfluid
Hilbert space. This is exactly the matrix element of the helium number
density operator. The corresponding experimental observable, the differential rate of neutron momentum and energy transfer into the superfluid, is known as the Dynamical Structure Function and defined as 
\begin{equation}
S(\boldsymbol{k},\omega)=\frac{m_\text{He}}{\rho}\sum_{\beta}\frac{|\langle\varphi_{\beta}(\boldsymbol{k},\omega)|n_{\boldsymbol{k}}|\Omega\rangle|^{2}}{|\langle\varphi_{\beta}(\boldsymbol{k},\omega)|\varphi_{\beta}(\boldsymbol{k},\omega)\rangle|^{2}}\delta(E_{i}-E_{f}-\omega),\label{eq:dynamicstructure}
\end{equation}
summed over all possible states $\beta$, and normalized by the average
number density $\rho/m_\text{He}$. The scattering rate is then given by
\begin{equation}
\Gamma_{\text{n}}=2\pi \frac{\rho}{m_\text{He}}\left(\frac{2\pi b_{n}}{m_{n}}\right)^{2}\int\frac{d^{3}k}{(2\pi)^{3}}S(\boldsymbol{k}, \omega).\label{eq:GammaSinglePh}
\end{equation}
In the limit where only a single quasiparticle is emitted, the degeneracy of the multi-quasiparticle spectrum is avoided and the final state phase space is greatly simplified. In this limit the DSF reduces to the static structure function (SSF) (see \cref{fig:ssf}):
\begin{equation}
    S(\boldsymbol{k},\omega)=S(k)\delta(\omega-\omega(k)),
    \label{eq:SSF}
\end{equation}
which for single phonon production takes the form 
\begin{equation}
S(k)=\frac{k}{2m_{\text{He}}c_{s}}.
\label{eq:Sofk}
\end{equation}
This simplification of the scattering rate in terms of the dynamic structure function applies to DM scattering as well, with the only difference being the coupling constant.

\subsubsection{Helium emitting quasiparticles}
\label{sec:he2quasi}
The neutron and DM scattering processes we have considered thus far are perturbative because the incident particle's kinetic energy is much greater than its interaction potential with helium atoms. In this section we consider another process: the emission of quasiparticles by energetic helium atoms. The superfluid is strongly coupled, as can be seen from the He-He scattering potential presented in \cref{sec:appendix}, so this process is non-perturbative and the procedure presented above cannot be applied. Instead, to study the emission rate, we introduce effective current-current couplings that are applicable over the quasiparticle momentum range.

\subsubsection*{Effective current-current couplings}

Since the superfluid system is strongly coupled, the perturbative matrix element calculation presented in previous sections cannot be applied directly to helium atoms within the fluid. To proceed, we treat the fast moving helium atom as a hard background particle and construct effective operators from the helium $U(1)$ currents expressed both in terms of the helium atoms and the phonon degrees of freedom. The scalar and vector components of the currents of the helium $\Phi$ and the phonon $\pi$ are respectively given by
\begin{align}
J^0_{\rm He} & =\Phi^{\dagger}\Phi,\quad 
 {J}^i_{\rm He}=\frac{-i}{2m_{\rm He}}\left(\Phi^{\dagger}\partial_i \Phi-\Phi\partial_i \Phi^{\dagger}\right),\label{eq:ClassicalJ}
\end{align} 
and
\begin{align}
J^{0} & =\frac{\sqrt{\rho}}{m_{\rm He} c_s} \dot{\pi}+\cdots,\quad 
{J}^i=-\frac{c_s \sqrt{\rho}}{m_{\rm He} }\partial_i\pi+\cdots.\label{eq:PhononJ}
\end{align}
Rotons and other quasiparticles are the high momentum excitations of phonons and can be viewed as the Goldstone modes associated with
the $U(1)$ symmetry as well. Therefore, we expect the matrix element of $J^\mu$, $\langle \psi | J^\mu | \Omega \rangle $, to be non-zero.
However, we cannot write the $J^\mu$ operator for quasiparticles as in \cref{eq:PhononJ}. Nevertheless, its matrix element can be related to the Dynamical~(Static) Structure Function. 
Note that the DSF only gives the value of 
$| \langle \varphi_\beta | J^0 | \Omega \rangle|^2$, but the $J^i$ matrix element can then be linked to the $J^0$ matrix element via current conservation $\partial_0 J^0 + \partial_i J^i = 0 $.

The simplest operators that can be formed from the two $U(1)$ currents are $J^0 J^0_{\rm He}$ and $J^i J^i_{\rm He}$.
Given the scaling dimensions of these operators $[J^0] = [J^i] = 3$, we can write down the effective Lagrangian as
\begin{equation}
    \mathcal{L}_{JJ} = \lambda_1 \frac{1}{m_{\rm He} \Lambda} J^0 J^0_{\rm He} +  \lambda_2  \frac{m_{\rm He} } {\Lambda^3} J^i J^i_{\rm He} \ ,
    \label{eq:Ljj}
\end{equation}
where $\lambda_1$ and $\lambda_2$ are undetermined ${\cal O}(1)$ coupling constants.
We take the $U(1)$ current $J^\mu_{\rm He}$ 
to be the helium current from \cref{eq:ClassicalJ} and the
current $J^\mu$ to be that of quasiparticles. To obtain the helium atom interactions with phonons, we substitute the explicit form of the currents from \cref{eq:ClassicalJ} and \cref{eq:PhononJ}:
\begin{align}
\mathcal{L}_{\text{He-ph}} &= \frac{1}{m_{\rm He} \Lambda} \, \frac{\sqrt{\rho}}{m_{\rm He} c_s} \left(\lambda_{1}\dot{\pi}\Phi^{\dagger}\Phi
    +\lambda_2\, c_s^2  \frac{m_{\rm He}^2 } {\Lambda^2} \,  \partial_i \pi\, \frac{i\Phi^{\dagger}\partial_i\Phi+h.c.}{2m_\text{He}}\right) \nonumber\\
&= \frac{1}{\sqrt{m_{\rm He} \Lambda} }  \left(\lambda_{1}^\prime \dot{\pi}\Phi^{\dagger}\Phi
    +\lambda_{2}^\prime \partial_i \pi\, \frac{i\Phi^{\dagger}\partial_i \Phi+h.c.}{2m_\text{He}}\right),
    \label{eq:ClPhJJ}
\end{align}
where we introduce the dimensionless couplings $\lambda_1^\prime=  \sqrt{ \frac{{\rho}}{m_{\rm He}^{3} c_s^2 \, \Lambda}} \, \lambda_1  $ 
and $\lambda_2^\prime=  \sqrt{ \frac{{\rho} m_{\rm He} c_S^2}{ \Lambda^5}} \, \lambda_2  $ to simplify the expressions for the phonon emission rate.

The single phonon emission rate can be derived by using \cref{eq:ClPhJJ}, or by using the Lagrangian in \cref{eq:Ljj} and the SSF. 
For higher momentum quasiparticles, including rotons, 
we can use the current-current coupling Lagrangian \cref{eq:Ljj} to cast the emission rate in terms of the DSF and
the couplings $\lambda_1$ and $\lambda_2$, which are expected to acquire scale dependence from the underlying strongly coupled dynamics. The exact behavior of the scale dependence is beyond the scope of this work, and for concreteness we will assume $\lambda_1$ and $\lambda_2$ to take effectively constant values over the relevant momentum range.

\subsubsection*{Helium-phonon emission rate}

In the interest of generality we shall adopt the formula \cref{eq:ClPhJJ} to calculate the
single phonon emission rate, leaving the coupling constants undetermined. The first operator in our Lagrangian has the same form as the DM-helium or neutron-helium coupling up to the presence of $\lambda_1$, while the second term contributes a factor of $\lambda_2 \boldsymbol{v}_\text{He} \cdot \boldsymbol{k}/\omega$ at the matrix element level. Here $\boldsymbol{v}_\text{He}$ is the hard helium velocity and $(\omega, \boldsymbol{k})$ is the energy and momentum of the phonon. Integrating over the allowed phonon momenta yields the emission rate:
\begin{align}
    \Gamma_{\text{ph}} &= \frac{c_{s}\pi}{m_\text{He}\Lambda}\int\frac{kd^{3}k}{(2\pi)^{3}}\left(\lambda_{1}^\prime+\lambda_{2}^\prime\frac{\boldsymbol{v}\cdot\boldsymbol{k}}{\omega(k)}\right)^{2}\delta(E_{i}-E_{f}-\omega) \nonumber\\
    & =(\lambda_{1}^\prime+\lambda_{2}^\prime)^2\frac{ k_{\max}^3 c_s }{12\pi\Lambda p_\text{He}},
\end{align}
where we have substituted the single phonon form of the SSF \cref{eq:Sofk} and $k_{\max} \simeq 1 ~\text{keV}$ is the largest valid momentum for the phonon mode. We have also assumed that the helium momentum is much larger than the phonon momenta: $p_\text{He}\gg k $.

\subsubsection*{Helium-quasiparticle emission rate}

The effective Lagrangian \cref{eq:Ljj} leads to the following order-of-magnitude estimate for the single quasiparticle emission rate
\begin{equation}
\Gamma = 2\pi \frac{\rho}{m_\text{He}}\int \frac{d^{3}k} {(2\pi)^{3}}\left(\frac{\lambda_1}{m_{\text{He}}\Lambda}
+ \frac{\lambda_2 m_{\text{He}}}{\Lambda^3} \frac{\boldsymbol{v}_{\text{He}}\cdot\boldsymbol{k}\omega}{k^2}\right)^{2}S(k)\delta(E_i -E_f -\omega),
\end{equation}
where we assume that in Fourier space $\tilde{\boldsymbol{J}}=\boldsymbol{k}\omega\tilde{J}^0 /k^2$, meaning that for quasiparticles the probability current is aligned with the direction of its momentum. Here we leave the integration as is, since, as mentioned above, the couplings $\lambda_{1}$ and $\lambda_2$ might be scale dependent. In the case of multiple quasiparticle emission the squared current matrix element is promoted from the SSF $S(k)\delta(E_{i}-E_{f}-\omega)$ to the DSF encompassing the complete superfluid response $S(\boldsymbol{q},\omega)$. 
To obtain the correct helium scattering rate, a novel helium scattering experiment might be needed to 
probe the ``Helium-quasiparticle" structure function. 
Note that we do not consider the ``absorption'' of these energetic helium atoms once they have lost sufficient energy. Such a process (where the free slow helium excitation is more appropriately described by a configuration of quasiparticles) may in principle contribute to the helium cascade, but is hard to treat due to the inherently nonperturbative dynamics linking degrees of freedom across different phases of helium.

%% file: process.tex
\section{Decay and scattering processes}
\label{sec:crosssecton}

After building the EFT framework in \cref{sec:theory}, in this section we apply the EFT method to derive 
the rates and cross sections for several relevant processes. While some of these results have appeared previously in the 
literature (see, e.g., \cite{landau1987statistical,landau1949theory,Nicolis_2018}), here we summarize them in one place 
for completeness (see \cref{tab:summary} in \cref{sec:summarycs} below), as an illustration of the EFT approach. 
In the following subsections we shall consider phonon decay and self-scattering, roton self-scattering and phonon-roton scattering. 
The rates of DM and helium atom emitting quasiparticles were already presented in \cref{sec:neutral2quasi}, and to avoid repeating, 
we shall not consider them again in this section (but still summarize in \cref{tab:summary}). Finally, quasiparticles with momenta above $4.5 ~\text{keV}$ may decay to two-roton final states, but this rate may only be roughly estimated by dimensional analysis.

These results can be applied to sub-GeV dark matter detection in the helium superfluid and to study of the superfluid itself. In particular, after the emission of quasiparticles from fast helium atoms, we must also consider the interactions of quasiparticles 
within the cascade.

\subsection{Phonon decay}

\begin{figure}[t]
\begin{center}
\includegraphics[scale=0.4]{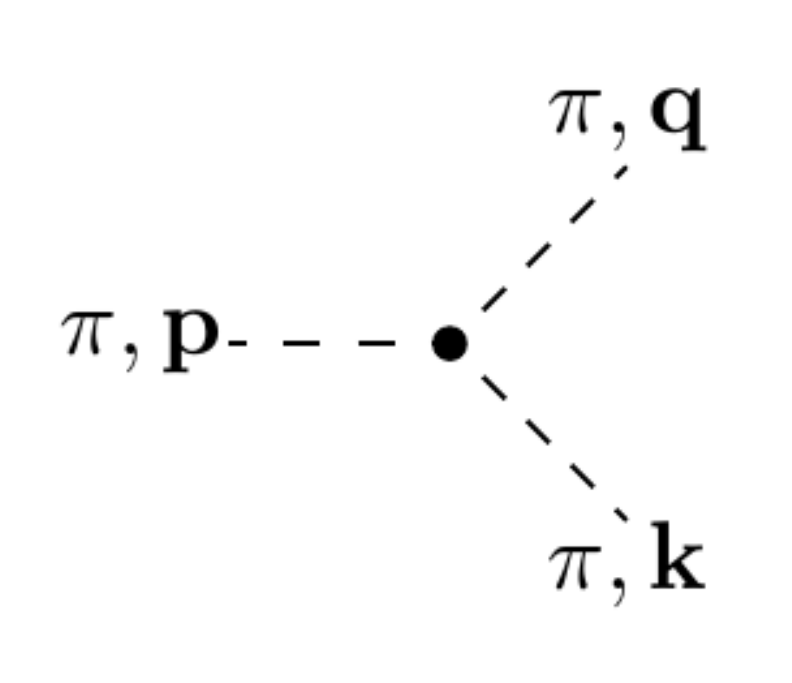}
\end{center}
\caption{The Feynman diagram and notation for the phonon decay process.}
\label{fig:decay}
\end{figure}

As depicted in \cref{fig:decay}, the process of phonon decay consists of one incoming and two outgoing phonon states, which are connected by the 3-point interaction terms in the Lagrangian in \cref{eq:Lphononinteraction}. The matrix element of this decay is given by the Feynman rule:
\begin{equation}
i\mathcal{M}=\frac{c_{s}^{5/2}}{\Lambda^{2}}\left[k\, \boldsymbol{p}\cdot\boldsymbol{q}+p\, \boldsymbol{q}\cdot\boldsymbol{k}+q\, \boldsymbol{p}\cdot\boldsymbol{k}+(2u-1)pqk\right] \, ,
\end{equation}
where $\boldsymbol{p}$ is the initial phonon momentum, $\boldsymbol{q}$ and $\boldsymbol{k}$ are the momenta of the daughter phonons  (see \cref{fig:decay}),
and $u$ is defined in \cref{eq:Pprimes}. 
The differential decay rate for this process is then written as
\begin{equation}
d\Gamma=\frac{1}{2E_{p}}|\mathcal{M}|^{2}\frac{d^{3}q}{(2\pi)^{3}2E_{q}}\frac{d^{3}k}{(2\pi)^{3}2E_{k}}(2\pi)^{4}\delta^{4}(p-q-k).
\end{equation}
Note that this is not a rate calculated in the phonon rest frame, but a lab frame calculation. As such this decay rate formula carries a factor of $1/(2E_p)$.

The final state phase space integration is somewhat nontrivial.
Because the dispersion is nearly linear, in order to conserve energy and momentum the daughter phonons must both move nearly parallel to the parent phonon's direction, and thus all three vectors $\boldsymbol{p},\boldsymbol{q},\boldsymbol{k}$
are approximately aligned (see \cite{baym2020searching} for details on the allowed kinematics in the presence of non-linear dispersion effects). The total rate of phonon decay is then 
\begin{equation}
\Gamma=\frac{(u+1)^{2}c_{s}p{}^{5}}{120\pi\Lambda^{4}}.
\end{equation}

\subsection{Phonon-phonon scattering}

\begin{figure}
\begin{center}
\includegraphics[scale=0.4]{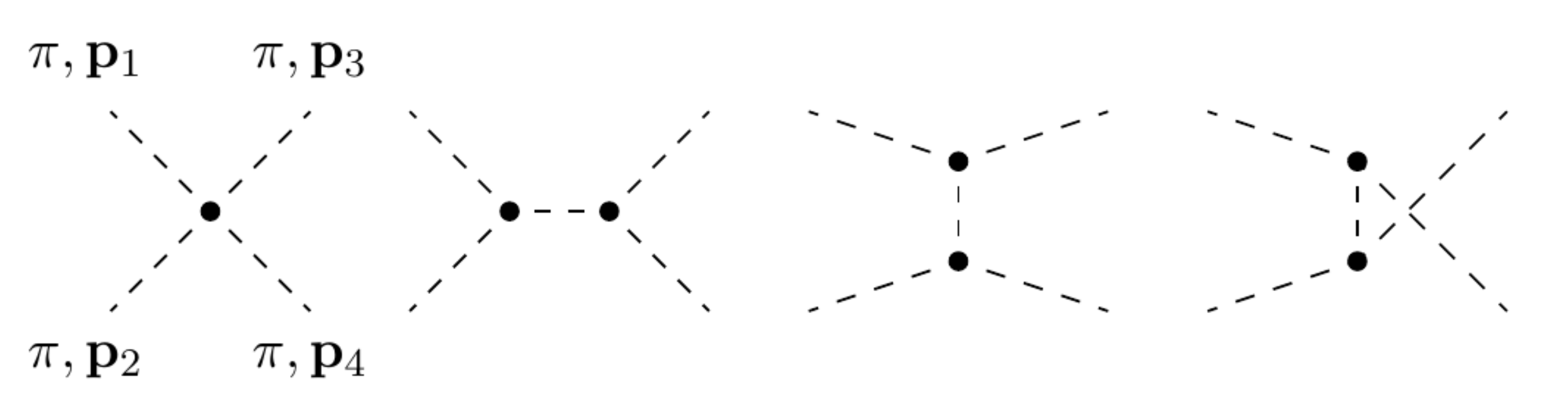}
\end{center}
\caption{Feynman diagrams and notation for the phonon-phonon scattering process. The black dots indicate 4-point or 3-point interaction terms in the Lagrangian in \cref{eq:Lphononinteraction}. }
\label{fig:phononsi}
\end{figure}

As shown in \cref{fig:phononsi}, phonon scattering may occur by the $s$, $t$, and $u$ channel diagrams and through the 4-point contact interaction. 
The first important observation is that the complete process is dominated by the $s$-channel diagram. To see this, consider the $s$-channel virtual phonon propagator 
\begin{equation}
\Delta_{F}(\boldsymbol{k},E)=\frac{i}{\left(E_{p_{1}}+E_{p_{2}}+E_{\boldsymbol{p}_{1}+\boldsymbol{p}_{2}}\right)\left(E_{p_{1}}+E_{p_{2}}-E_{\boldsymbol{p}_{1}+\boldsymbol{p}_{2}}\right)}.
\end{equation}
The second factor in the denominator is proportional to $\vert\boldsymbol{p}_{1} \vert +\vert \boldsymbol{p}_{2}\vert -|\boldsymbol{p}_{1}+\boldsymbol{p}_{2}|$, which vanishes when $\boldsymbol{p}_{1}\parallel\boldsymbol{p}_{2}$. Only the s-channel tree diagram has this property, which exhibits a co-linear divergence of the initial phonons.

The second relevant observation is that the phase space support for the $s$-channel diagram is dominated by the collinear limit when $\vert \boldsymbol{p}_1 \vert \gg \vert \boldsymbol{p}_2 \vert$, i.e., ``hard-soft" scattering. In contrast, for scattering between phonons with similar momentum, the constrained kinematics from energy and momentum conservation allows for only forward scattering and is therefore suppressed by phase space.
For hard-soft scattering, the differential cross-section is largest at the smallest allowed value of the angle $\theta$ between $\boldsymbol{p}_1$ and $\boldsymbol{p}_2$. The nonlinear dispersion introduces a cutoff $\theta^{2}\ge6\frac{\gamma}{\Lambda^{2}}p_{1}p_{3}p_{4}/p_{2}$. As a result, the cross section times the relative velocity $\Delta v$ between the two initial phonons, averaged over $\theta$, has the following expression 

\begin{align}
\langle \sigma \Delta v\rangle_{\theta} & \simeq\frac{(2u+2)^{2}c_{s}p_{1}^{4}}{96^{2}\gamma\pi\Lambda^{6}},
\end{align}
which is the dominant contribution from
the co-linear divergence in the $s$-channel diagram. We see that the result is proportional to
$p_{1}^{4}/\Lambda^{6}$, where $p_{1}$ is the hard phonon momentum.
The other diagrams (as well as the $s$-channel diagram away from the ``hard-soft" limit) produce subleading terms of order $p^{6}/\Lambda^{8}$ (or higher),
where $p$ is a generic initial state momentum.

\subsection{Roton-roton scattering}

\begin{figure}
\begin{center}
\includegraphics[scale=0.3]{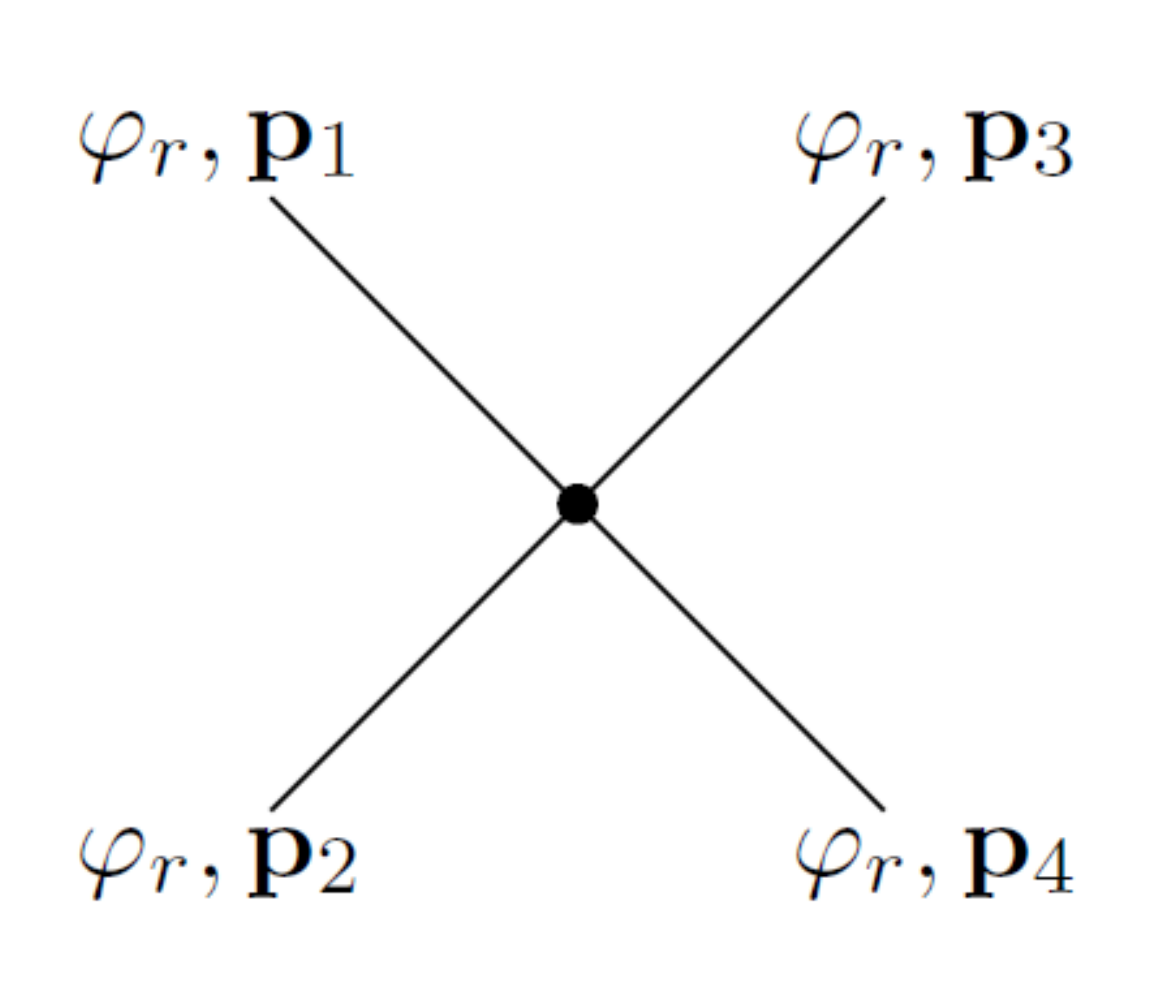}
\end{center}
\caption{Feynman diagram and notation for the roton-roton scattering process.}
\label{fig:rotonrotonscatter}
\end{figure}

The roton-roton scattering process is depicted in \cref{fig:rotonrotonscatter}, where the black dot stands for the single marginal interaction term in the roton-roton self interaction Lagrangian \cref{eq:rotoninteractions}. We shall naively take the 4-roton coupling constant $\lambda_r$ to be independent of momenta.
While the matrix element is simply $i\lambda_{r}/m_{\ast}p_{\ast}$, the highly nonlinear roton dispersion relation means that the final state phase space requires careful treatment.
The dominant contribution is from rotons with momenta close to $p_{\ast}$. After a clever change of variables during phase space integration \cite{landau1949theory}, the total cross section is given by

\begin{equation}
\sigma=
\frac{2\lambda_{r}^{2}}{|\boldsymbol{v}_{1}-\boldsymbol{v}_{2}|\cos\frac{\theta}{2}m_{\ast}p_{\ast}},
\end{equation}
where $|\boldsymbol{v}_{1}-\boldsymbol{v}_{2}|$ is the relative group
velocity of the rotons and, as before, $\theta$ is the angle between $\boldsymbol{p}_1 $ and $\boldsymbol{p}_2 $. The experimental value of $\lambda_r/m_\ast p_\ast=4.1\times10^{5}\ \text{GeV}^{-2}$ then implies 
that $\lambda_{r}=0.93$.

\subsection{Roton-phonon scattering}

\begin{figure}[tbp]
\begin{center}
\includegraphics[scale=0.4]{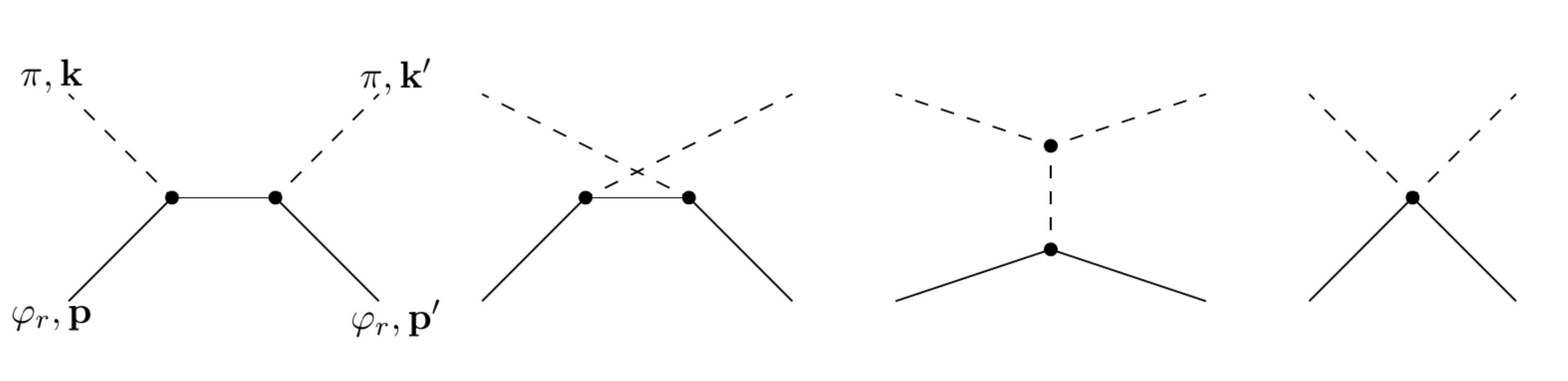}
\end{center}
\caption{Feynman diagram and notation for the roton-phonon scattering process.}
\label{fig:rotonphonondiags}
\end{figure}

Consider the roton-phonon scattering process depicted in \cref{fig:rotonphonondiags}, where the roton momentum $\mathbf{p}$ is taken to be much harder than the incident phonon momentum $\mathbf{k}$, i.e., $p \gg k$.
The leading contributions to the matrix element arise from the $s$-channel, $u$-channel and 4-point diagram from the contribution of $y_3^\prime$, $y_4$ and $y_4^{\prime\prime}$ terms in \cref{eq:rotonphononinteractionC}. It has the following expression\footnote{
Due to different derivation approaches, ref.~\cite{Nicolis_2018} has an additional contribution to the matrix element \cref{eq:PhoRoM}, 
$
\frac{i c_s}{m_\ast\Lambda^4}\left[\left(\boldsymbol{k}\cdot\boldsymbol{p}\right)^{2}+\left(\boldsymbol{k}^{\prime}\cdot\boldsymbol{p}\right)^{2}\right].
$
}:
\begin{equation}
i\mathcal{M} =-\frac{ic_{s}^{2}}{\Lambda^{4}}\biggl\{\left(\boldsymbol{k}\cdot\boldsymbol{k}^{\prime}\right)\left(\boldsymbol{k}+\boldsymbol{k}^{\prime}\right)\cdot\frac{\boldsymbol{p}}{k}+\frac{\left(\boldsymbol{k}\cdot\boldsymbol{p}\right)^{2}\left(\boldsymbol{k}^{\prime}\cdot\boldsymbol{p}\right)^{2}}{k^{2}p^{2}m_{\ast}c_{s}}+k^{2}\Lambda y_{4}
\biggr\}.
  \label{eq:PhoRoM}
\end{equation}
The differential cross section is
\begin{align}
d\sigma & =\frac{d^{3}p'}{(2\pi)^{3}}\frac{d^{3}k'}{(2\pi)^{3}2c_{s}k'}\frac{|\mathcal{M}|^{2}}{2c_{s}^{2}k}(2\pi)^{4}\delta^{4}(p+k-p'-k')=\frac{|\mathcal{M}|^{2}}{16\pi^{2}c_{s}^{4}}d\Omega,
\end{align}
and the total cross section is given by
\begin{equation}
\langle\sigma\rangle_{\Omega}=\frac{1}{4\pi}\left[\frac{1}{25\Lambda^{8}}\frac{p_{\ast}^{4}k^{4}}{m_{\ast}^{2}c_{s}^{2}}+\frac{2432+45\pi^{2}}{11520\Lambda^{8}}p_{\ast}^{2}k^{4}+\frac{2y_{4}}{9\Lambda^{7}}\frac{p_{\ast}^{2}k^{4}}{m_{\ast}c_{s}}+\frac{y_{4}^{2}}{\Lambda^{6}}k^{4}\right],
\label{eq:PhoRoSigma}
\end{equation}
in which the cross section $\sigma$ is a function of the spatial angle $\Omega$ between initial phonon and roton. For simplicity, here we have taken the average over $\Omega$.

\begin{table}[htp!]
\centering{}\centering{ %
\begin{tabular}{|>{\centering}m{2.8cm}|>{\centering}m{2.2cm}|>{\centering}m{8.5cm}|}
\hline 
Process  & Diagram  & Result \tabularnewline
\hline 
\vspace{0.3in}
\begin{centering}
Phonon decay 
\par\end{centering}
\vspace{-0.2in}
\centering{} 
\[
\pi\to\pi+\pi
\]
 & \includegraphics[scale=0.2]{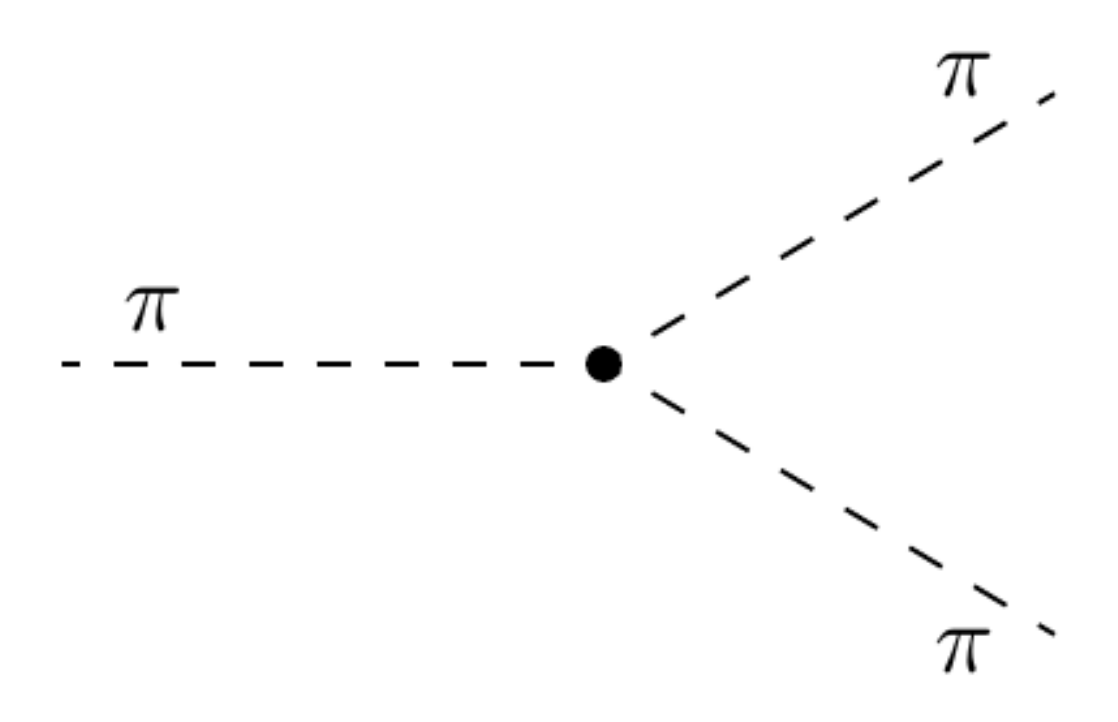}  & \centering{} 
\[
\Gamma=\frac{(u+1)^{2}c_{s}p{}^{5}}{120\pi\Lambda^{4}}
\]
\tabularnewline
\hline 
\vspace{0.3in}
\begin{centering}
Phonon self-scattering 
\par\end{centering}
\vspace{-0.2in}
\[
\pi+\pi\to\pi+\pi
\]
 & \includegraphics[scale=0.2]{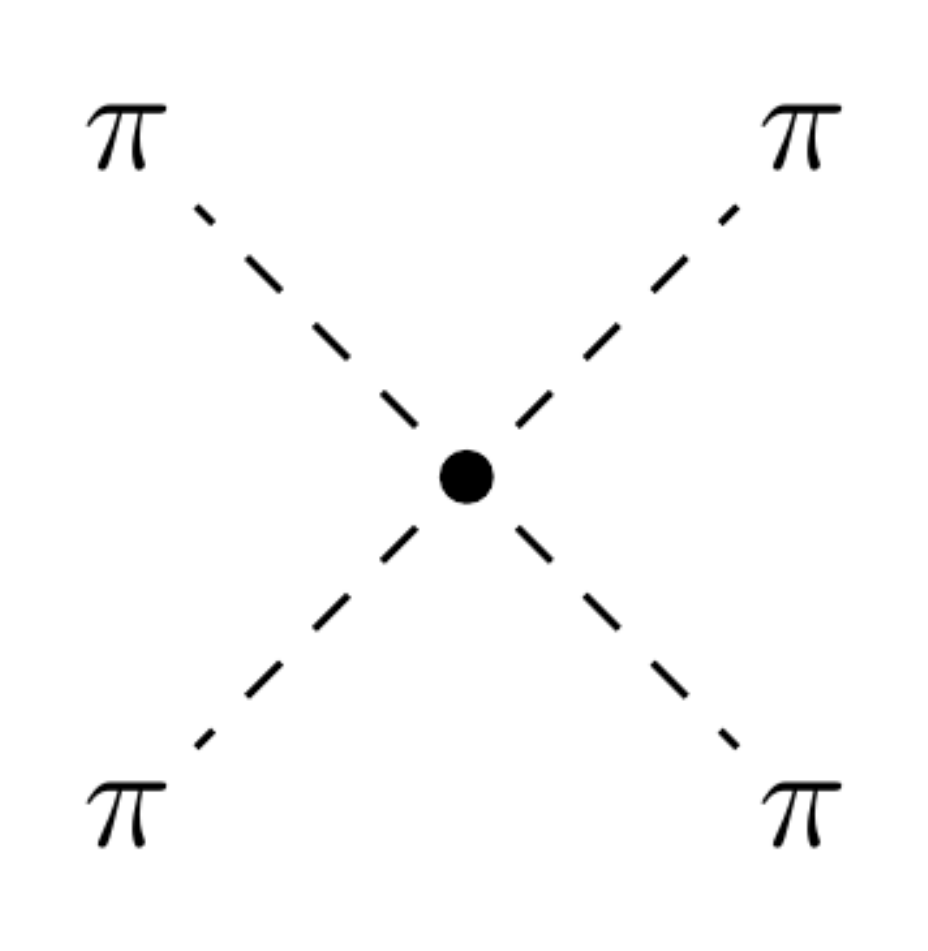}  & \centering{} 
\[
\langle\sigma \Delta v\rangle_{\theta}\simeq\frac{(2u+2)^{2}c_{s}p_{1}^{4}}{96^{2}\gamma\pi\Lambda^{6}}
\]
\tabularnewline
\hline 
\vspace{0.3in}
Roton self-scattering
\vspace{-0.2in}
\[
\varphi_{r}+\varphi_{r}\to\varphi_{r}+\varphi_{r}
\]
 & \includegraphics[scale=0.2]{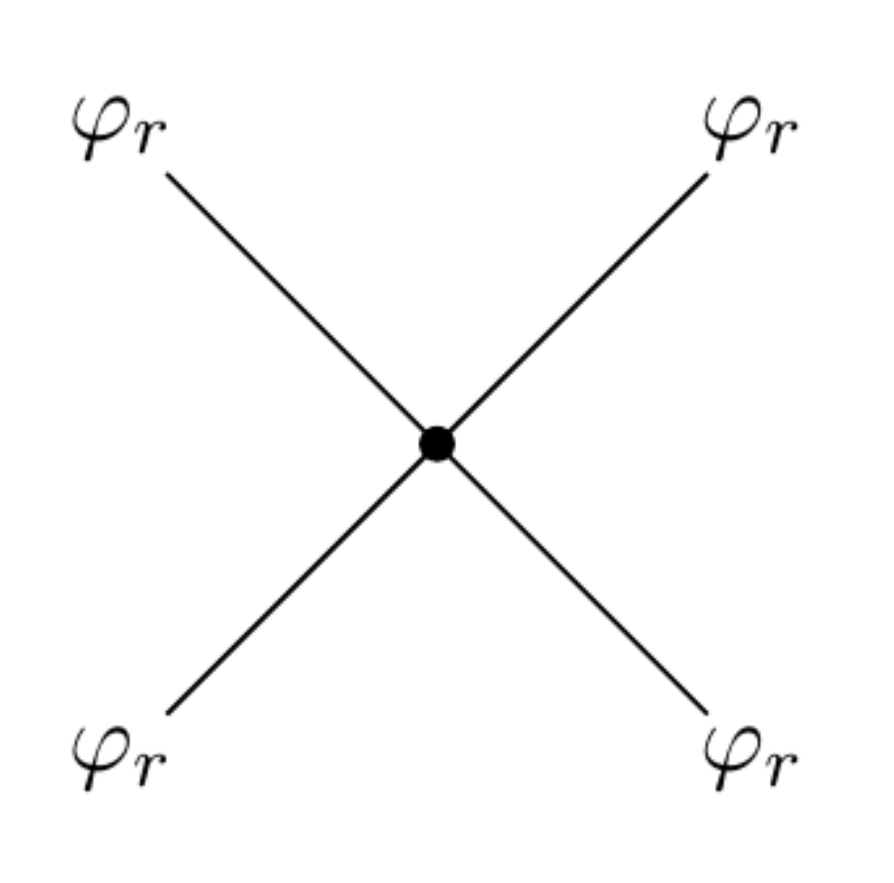}  & \centering{} 
\[
\sigma=\frac{2\lambda_{r}^{2}}{|\boldsymbol{v}_{1}-\boldsymbol{v}_{2}|\cos\frac{\theta}{2}m_{\ast}p_{\ast}}
\]

\tabularnewline
\hline 
\vspace{0.3in}
\begin{centering}
Roton-phonon scattering 
\par\end{centering}
\vspace{-0.2in}
\[
\pi+\varphi_{r}\to\pi+\varphi_{r}
\]
 & \includegraphics[scale=0.2]{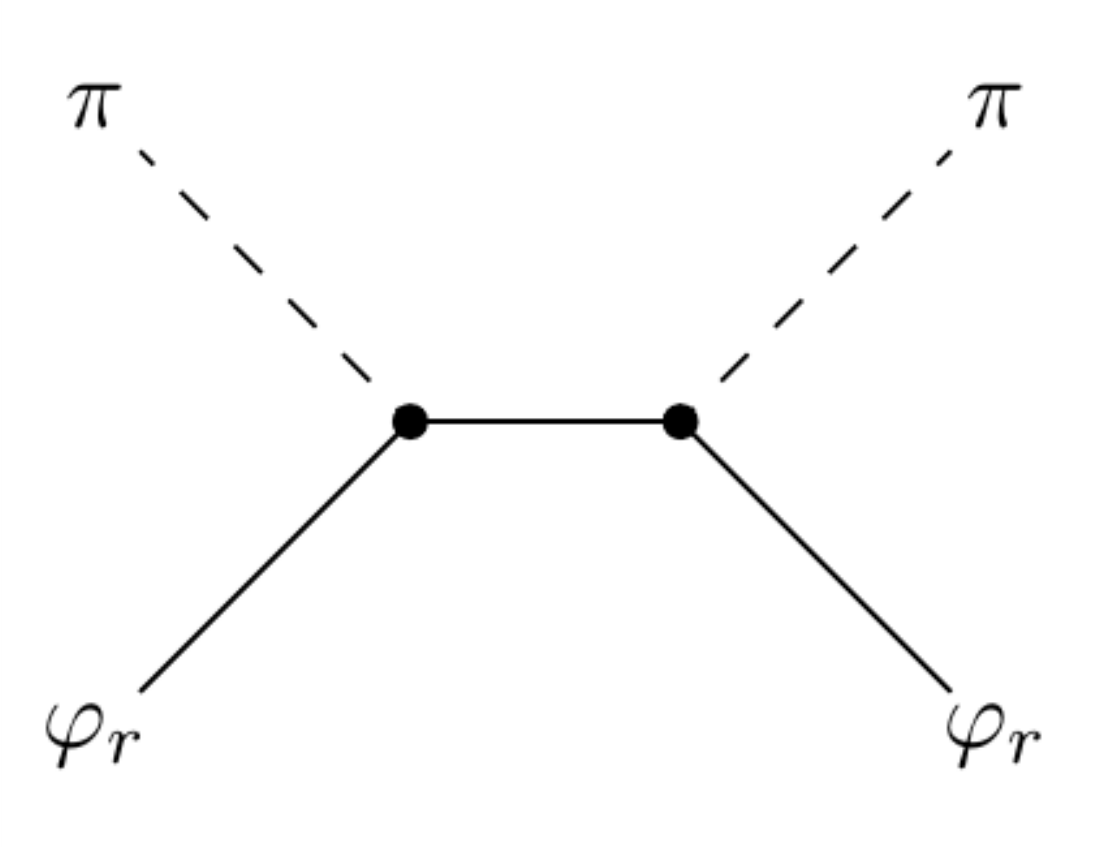}  & \centering{} 
\begin{eqnarray*}
\langle\sigma\rangle_{\Omega} &=&\frac{1}{4\pi}\biggl[\frac{1}{25\Lambda^{8}}\frac{p_{\ast}^{4}k^{4}}{m_{\ast}^{2}c_{s}^{2}}+\frac{2432+45\pi^{2}}{11520\Lambda^{8}}p_{\ast}^{2}k^{4}\\
&+&\frac{2y_{4}}{9\Lambda^{7}}\frac{p_{\ast}^{2}k^{4}}{m_{\ast}c_{s}}+\frac{y_{4}^{2}}{\Lambda^{6}}k^{4}\biggr].
\end{eqnarray*}

\tabularnewline
\hline 
\vspace{0.3in}
\begin{centering}
Helium emits quasiparticles
\par\end{centering}
\vspace{-0.2in}
\[
\text{He}\to\text{He}+\varphi
\]
 & \includegraphics[scale=0.2]{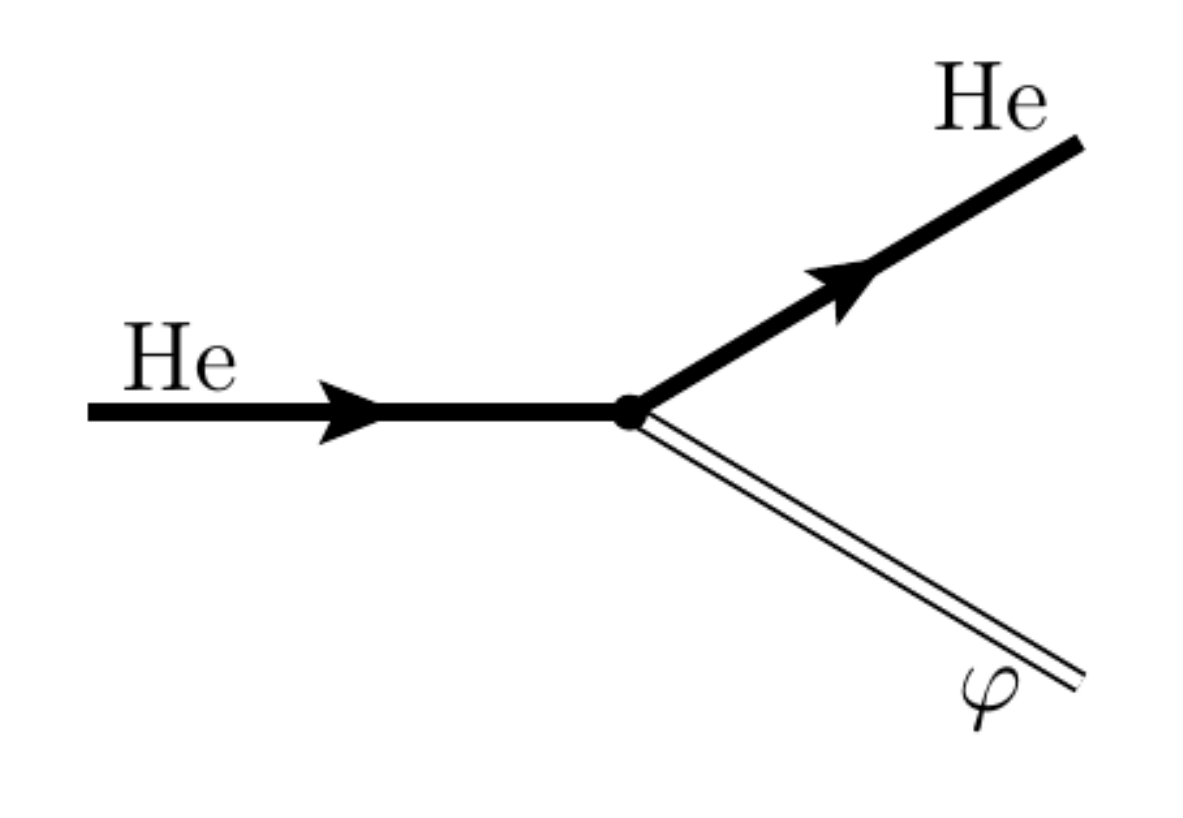}  & \centering{} 
\[
    \footnotesize{\Gamma_m =  \frac{2\pi\rho}{m_\text{He}} \int \frac{d^{3}k} {(2\pi)^{3}}\left(\frac{\lambda_1}{m_{\text{He}}\Lambda}
+ \frac{\lambda_2 m_{\text{He}}}{\Lambda^3} \frac{\boldsymbol{v}_{\text{He}}\cdot\boldsymbol{k}\omega}{k^2}\right)^{2}S(\boldsymbol{k},\omega)}
\]

\vspace{-0.2in}

\[
\text{Single emission:} \, S(\boldsymbol{k},\omega)\to S(k)\delta(E_i-E_f-\omega)
\]

\tabularnewline
\hline 
\vspace{0.3in}
\begin{centering}
DM emits quasiparticles 
\par\end{centering}
\vspace{-0.2in}
\centering{} 
\[
\text{DM}\to\text{DM}+\varphi
\]
 & \includegraphics[scale=0.18]{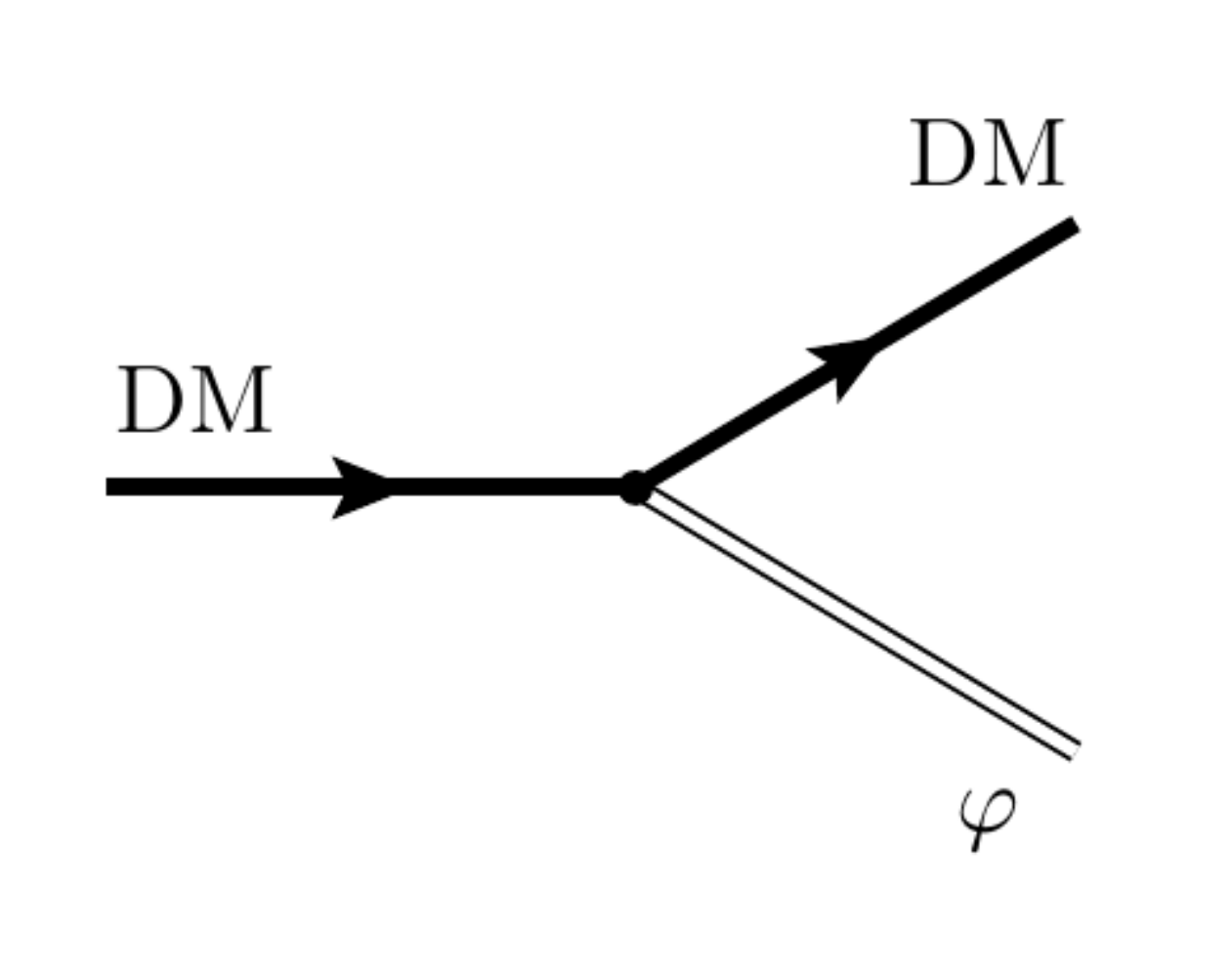}  & \centering{} 

\[
\Gamma_s=\frac{\rho}{2\pi m_{\text{He}}v_{\text{DM}}}\left(\frac{\lambda}{m_{\text{\ensuremath{\sigma}}}^{2}}\right)^{2}\int k\,S(k)\,dk
\]
\[
\Gamma_m=\frac{2\pi\rho}{m_{\text{He}}}\left(\frac{\lambda}{m_{\text{\ensuremath{\sigma}}}^{2}}\right)^{2}\int \frac{d^{3}k}{(2\pi)^{3}}\,S(\boldsymbol{k},\omega)
\]
\tabularnewline
\hline 
\end{tabular}} \caption{\label{tab:summary} Summary of the main results from \cref{sec:crosssecton}. 
The UV parameter $\Lambda$ has the following conversion $\Lambda^4=\rho c_s$. $\Gamma_m$ and $\Gamma_s$ denote the multiple and single quasiparticle emission rate, respectively. }

\end{table}

\subsection{Summary}
\label{sec:summarycs} 

In \cref{tab:summary} we summarize the relevant results for the processes considered in the last two sections. The first column identifies the process, the second column lists a representative Feynman diagram, while the analytical result for the relevant observable quantity is given in the third column. 

There are some details of these calculations which merit brief discussion: 
\begin{itemize}
\item When a phonon lies in the linear dispersion regime, its decay daughters' momenta are colinear.
\item The phonon--phonon scattering rate is dominated by the hard-soft case, where one phonon has momentum much larger than the other. 
\item The roton coupling constant has been measured, $\lambda_r = 0.93$. 
\item In presenting the roton-phonon cross section we have included the dimensionless $y_4$ from \cref{eq:rotonphononinteractionC} and taken the approximation $\partial{p_\ast}/\partial{\rho}=p_\ast/\rho$ motivated by experiment.
\end{itemize}

%% file: conclusions.tex

In this work we used several EFT methods to construct free and
interacting Lagrangians for quasiparticles, neutral probes and helium atoms in superfluid $^{4}{\rm He}$. 
Then we calculated explicitly the dominant contributions to the scattering cross sections, decay rates and production rates involving these 
degrees of freedom.

We reviewed the interactions for \textbf{phonons} as Goldstone bosons of a spontaneously broken helium $U(1)$ symmetry. 
Their interactions are obtained by exploiting this $U(1)$ symmetry of the quantum action and expanding the effective Lagrangian 
about its equilibrium chemical potential. The expansion coefficients are obtained from thermodynamic measurements.

\textbf{Rotons} are special quasiparticles which lie on a local energy minimum of the dispersion curve. 
After expanding around the minimum, we used EFT power counting to show that the $\varphi_{r}^{4}$
term is the dominant marginal term in the roton self-interaction Lagrangian.

\textbf{Quasiparticles} (including {\bf rotons}) with momenta above $1$ keV can be treated as hard background particles when scattering with soft phonons.
We construct the interaction Lagrangian from power counting arguments, and from the theory of impurity particles interacting 
with a moving fluid. The latter method provides analytical expressions for the quasiparticle-phonon interaction coefficients.

We study the emission of quasiparticles from probe particles like {\bf DM} or {\bf neutrons}. The emission of phonons from DM and neutrons 
can be understood via the EFT approach. On the other hand, experimental results on neutron scattering off the superfluid provide additional information in terms of the dynamical structure function. By linking the DSF to the matrix element of the current $J^0$, we extended the EFT method to obtain the emission rates for quasiparticles as well.

A fast \textbf{helium atom} will emit quasiparticles during its strong non-perturbative interaction 
with the surrounding helium atoms in the superfluid. We proposed an effective current-current interaction Lagrangian 
to understand these processes. By utilizing the DSF and current conservation, we obtained the emission rates up to ${\cal O}(1)$ coefficients.

With the addition of the discussion  in \cref{sec:appendix} of helium atomic scattering, we have treated all relevant processes 
occurring after a sub-GeV dark matter particle scatters from a single helium atom in the superfluid $^{4}{\rm He}$. 
We leave the corresponding numerical simulation and experimental sensitivity projections to the future work \cite{MSXY}.
By constructing the EFT for quasiparticle interactions, we can additionally understand the 
processes of diffusion and thermalization of the produced quasiparticles in the superfluid ${}^4$He.

%% file: appendix.tex
\section{Helium-helium scattering}
\label{sec:appendix}

The scattering of a helium atom with momentum in the keV-MeV range off a stationary helium atom is
a non-perturbative process, since the atom's energy is comparable to
the interaction potential \cite{bennewitz1972he,feltgen1973determination,bishop1977low}.
We may obtain the phase shifts and thus the scattering cross section using a partial wave expansion of the Schrodinger equation.  Consider a wavefunction $\Psi(\boldsymbol{r})$
as a function of the relative distance vector $\boldsymbol{r}$ between the
two helium atoms. The incoming and outgoing asymptotic boundary condition
constrain the wavefunction as follows:

\begin{equation}
\Psi(\boldsymbol{r})|_{r\to\infty}=\exp(ikz)+f(\theta,k)\frac{\exp(ikr)}{r}.
\label{wavef}
\end{equation}
The first term is a plane incoming wave and the second term is the
outgoing spherical wave characterized by a probability distribution
$f(\theta,k)$. Here $k$ is the reduced momentum,
which is half of the incoming helium atom momentum in the
lab frame.

\begin{figure}[t]
  \centering
  \includegraphics[width=0.7\textwidth]{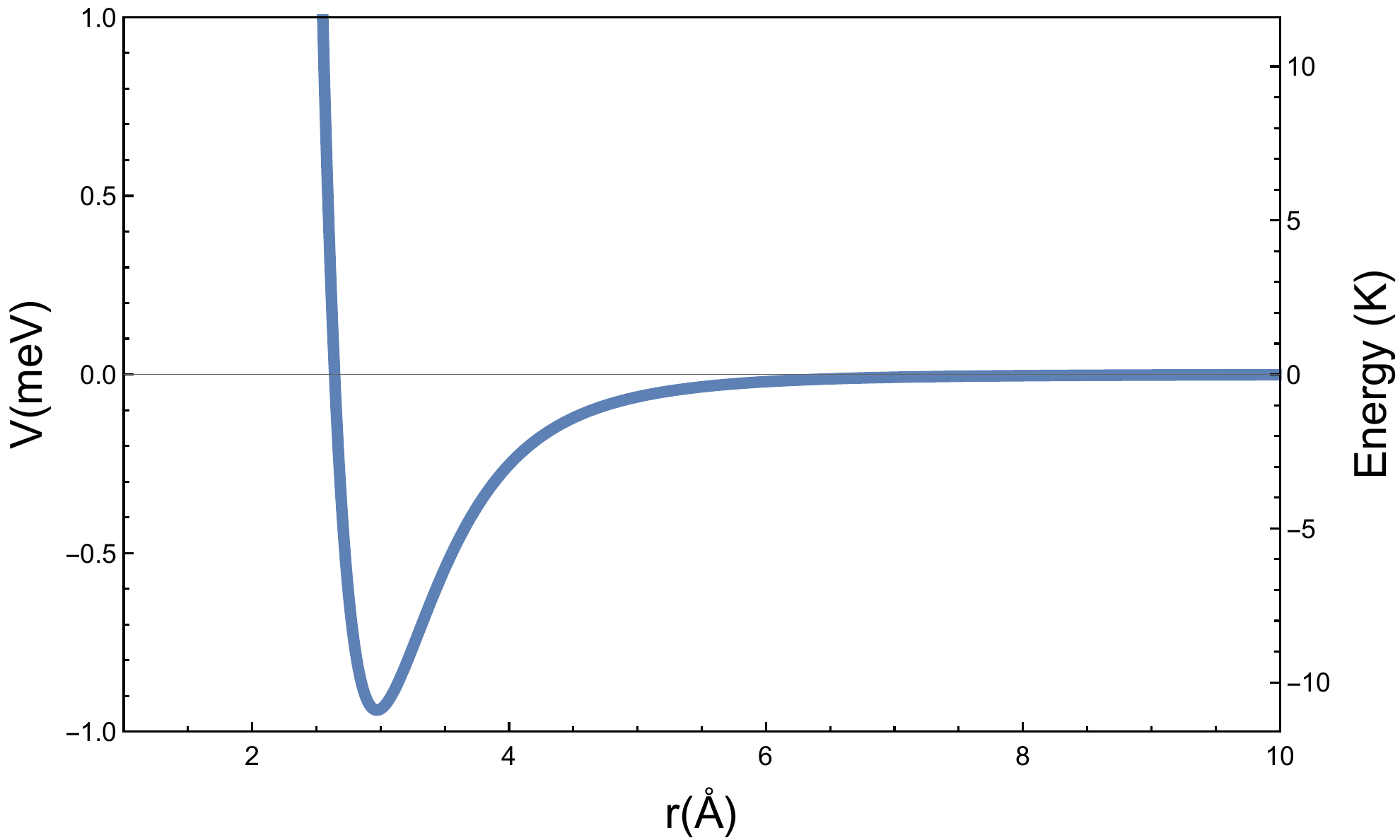}
    \caption{Helium-helium atomic potential fitted from the experimental results in \cite{feltgen1982unique}.}
  \label{pic:V(r)}
\end{figure}

To solve the scattering probability, we expand the asymptotic wavefunction as

\begin{equation}
\Psi(\boldsymbol{r})|_{r\to\infty}=\sum_{l=0}^{\infty}(2l+1)P_{l}(\cos\theta)\psi_{l}(r),
\label{Psiexpansion}
\end{equation}
where $\theta$ is the polar angle of the position vector $\boldsymbol{r}$
in the scattering plane. The radial degree of freedom $\psi_{l}(r)$
contains both the incoming and outgoing waves in \cref{wavef}, and satisfies
the radial Schrodinger equation:

\begin{equation}
\left(\frac{d^{2}}{dr^{2}}+k^{2}-2\mu V(r)-\frac{l(l+1)}{r^{2}}\right)\psi_{l}(r)=0,
\end{equation}
in which $V(r)$ is the potential between helium atoms depicted in \cref{pic:V(r)}. %
The solution
of $\psi_{l}(r)$ contains a phase shift $\delta_{l}(k)$ that depends
only on the helium momentum. 
The scattering amplitude is as follows:

\begin{equation}
f(\theta,k)=\frac{1}{2ik}\sum_{l=0}^{\infty}(2l+1)\left[\exp(2i\delta_{l})-1\right]P_{l}(\cos\theta).
\end{equation}
In the case of identical bosons, there is
an extra factor of 2 and the sum runs only over even $l$ values. The differential cross section is $|f(\theta,k)|^{2}$
and the total cross section is as follows:

\begin{equation}
\sigma_{\text{tot}}(k)=\frac{8\pi}{k^{2}}\sum_{l\in\text{even}}(2l+1)\sin^{2}\delta_{l}(k).
\end{equation}

The phase variation method
\cite{calogero1967variable} produces a first order differential equation
with respect to $r$ for a radial dependent phase shift $\delta_{l}(k,r)$ as:

\begin{equation}
\delta_{l}'(k,r)=-\frac{2\mu}{k}V(r)\left[kr\cos\delta_{l}(k,r)j_{l}(kr)-kr\sin\delta_{l}(k,r)y_{l}(kr)\right]^{2},
\end{equation}
where $j_{l}$ and $y_{l}$ are spherical Bessel functions. Using
an initial condition $\delta(0)\to0$, the equation can be numerically integrated in $r$, and the phase shift $\delta_{l}(k)$ is the asymptotic value
$\delta_{l}(k,\infty)$. 

The phase shifts may also be found using
the WKB approximation
\cite{miller1969wkb,miller1971additional}

\begin{equation}
\delta_{l}(k)=\lim_{r\to\infty}\left[\int_{r_{t}}^{r}k_{l}(r)dr-kr+(l+\frac{1}{2})\frac{\pi}{2}\right],
\end{equation}
where $r_{t}$ is the turning point corresponding to the classically
forbidden boundary, i.e., the point where the kinetic and potential energies are equal. The $k_{l}(r)$ is called the Langer-corrected wave-number

\begin{equation}
k_{l}(r)=\left[k^{2}-2\mu V(r)-\frac{l(l+1)}{r^{2}}\right]^{1/2}.
\end{equation}

\begin{figure}[tbp]
  \centering
  \includegraphics[width=0.7\textwidth]{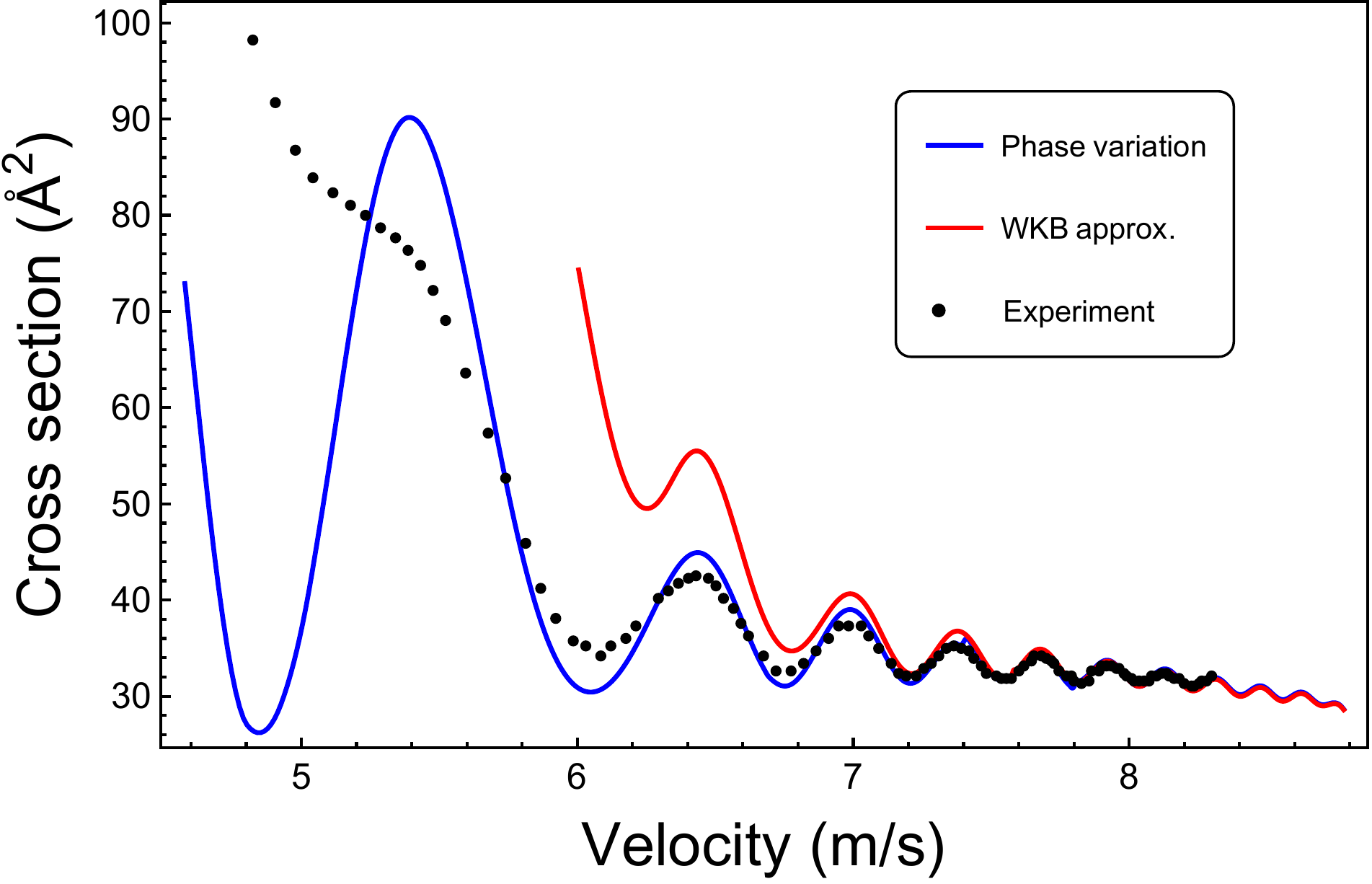}
    \caption{The $^{4}\text{He}-{}^{4}\text{He}$ scattering cross section as a function of the velocity of the initial helium beam in the lab frame. The experimental values (the black circles) are taken from \cite{feltgen1973determination}. We overlay the results from the two theoretical calculations discussed in the text, using the potential from \cref{pic:V(r)} --- the phase variation method (blue solid line) and the WKB method (orange solid line).}
    \label{fig:hehexsec}
\end{figure}

In \cref{fig:hehexsec} we show results for the $^{4}\text{He}- ^{4}\text{He}$ scattering cross section as a function of the velocity of the initial helium beam in the lab frame. The experimental values (the black circles) are taken from \cite{feltgen1973determination}. We overlay the results from the two theoretical calculations above, using the potential from \cref{pic:V(r)} --- the phase variation method (blue solid line) and the WKB method (orange solid line).
Note that the WKB approximation is only accurate at large momentum $k$, where it provides a reasonable fit to the data. 
We observe that the phase variation method agrees with the experimental data better than WKB, although it still deviates at very low momenta, due to the coupling between the helium atoms becoming strong.